\documentclass[notitlepage,twocolumn,pre,tightenlines,nofootinbib,showpacs,superscriptaddress,longbibliography]{revtex4-1}
\pdfoutput=1
\usepackage{amsmath}
\usepackage{amssymb,amsfonts,latexsym}
\usepackage{graphicx}
\usepackage[english,french]{babel}

\usepackage{color}
\usepackage{textcomp}
\usepackage[latin1]{inputenc}

\definecolor{blue}{rgb}{0,0,1}
\definecolor{bleu}{rgb}{0,0,0.8}
\definecolor{bleuf}{rgb}{0,0,0.9}
\definecolor{rougef}{rgb}{0.9,0,0}
\definecolor{green}{rgb}{0,0.5,0}
\definecolor{vert}{rgb}{0,0.8,0}
\definecolor{red}{rgb}{1,0,0}
\definecolor{pink}{rgb}{0.9,0.3,0.7}
\definecolor{azur}{rgb}{0,0.5,0.5}
\definecolor{orange}{rgb}{1,0.5,0.2}
\definecolor{brown}{rgb}{0.5,0,0}

\newcommand{\resub}[1]{#1}

\newcommand{\be}{\begin{equation}}
\newcommand{\ee}{\end{equation}}
\newcommand{\ben}{\begin{equation*}}
\newcommand{\een}{\end{equation*}}
\newcommand{\ba}{\begin{eqnarray}}
\newcommand{\ea}{\end{eqnarray}}

\newcommand{\leg}[1]{\textbf{#1}}

\begin{document}
%\graphicspath{{./Figures/}}

\title{Elasticité des empilements granulaires proche de la transition de blocage.\\
\emph{Elasticity of granular packings close to Jamming.}
}
\author{C. Coulais}
\affiliation{SPHYNX/SPEC, CEA-Saclay, URA 2464 CNRS, 91191 Gif-sur-Yvette, France}
\affiliation{Universit\'e Paris-Sud, CNRS, Lab FAST, Bat 502, Campus Universit\'e, Orsay, F-91405, France}
\affiliation{Huygens-Kamerlingh Onnes Lab, Universiteit Leiden, PObox 9504, 2300 RA Leiden, The Netherlands}
\author{A. Seguin}
\affiliation{SPHYNX/SPEC, CEA-Saclay, URA 2464 CNRS, 91191 Gif-sur-Yvette, France}
\affiliation{Universit\'e Paris-Sud, CNRS, Lab FAST, Bat 502, Campus Universit\'e, Orsay, F-91405, France}
\author{O. Dauchot}
\affiliation{EC2M, ESPCI-ParisTech, UMR Gulliver 7083 CNRS, 75005 Paris, France}

\begin{abstract}
Nous nous intéressons à la réponse mécanique au cisaillement d'un milieu granulaire modèle 
bidimensionel au-delà de la transition de blocage appelée aussi 
"jamming". Tout d'abord, nous développons le dispositif expérimental et nous 
combinons des techniques de suivi de particules et de photoelasticté afin de 
mesurer l'état de déformation et l'état de contrainte à l'échelle du grain. 
Ensuite, nous mettons en place un intrus capable d'extension radiale (un 
"ballon" 2D) afin de pouvoir cisailler l'empilement granulaire tout en 
conservant une géométrie axisymétrique. Nous sondons l'apparition des 
contraintes à l'échelle du grain pour des amplitudes de déformation inférieures à 
$10^{-2}$ et pour une gamme de fraction volumique évoluant de  $2\%$ de part et 
d'autre de la transition de blocage. Nous montrons ainsi que cette réponse 
mécanique induit des contraintes de cisaillement mais aussi des contraintes 
normales. De plus, nous identifions un régime élastique où les contraintes 
normales et les contraintes de cisaillement évoluent non-linéairement avec la 
déformation de cisaillement. Enfin, nous explicitons la relation entre 
l'apparition de cette non-linéarité et la transition de blocage et nous 
déterminons les relations constitutives de l'empilement.\\
\small{\emph{
We investigate experimentally the mechanical response to shear of a 2D packing of grains across the
jamming transition. First, we develop a dedicated experimental setup, together with tracking and photoelastic techniques in order
to prepare the packing in a controlled fashion and to quantify the stress and strain tensors at the grain scale. Second,
we install a inflating probe (a 2D "balloon"), which shears the packing with a cylindrical symmetry. We probe
experimentally stresses and strains for strain amplitudes as low as $10^{-3}$ and for a range of packing
fractions within $2\%$ variation around the jamming transition. We observe not only that shear strain
induces shear stresses, but also normal stresses. Moreover, we show that both shear and normal stresses behave nonlinearly with
the shear strain. Finally, we show by scaling analysis that the constitutive laws are controlled by the Jamming transition.}}
\end{abstract}


\maketitle

%\tableofcontents

%-----------------------Introduction-------------------------------------------
\section{Introduction}
Comprendre les propriétés mécaniques des empilements denses de particules 
athermiques telles que les grains, les mousses ou bien encore les émulsions 
demeure un défi aussi conceptuel que pratique. Les fluctuations, le désordre et 
l'anisotropie contrôlent la mécanique de ces systèmes et contrecarrent la plupart 
des tentatives de détermination des lois constitutives de ces matériaux. Un 
progrès considérable a été effectué lors de l'introduction d'un modèle simple de 
sphères molles non frottantes, contenant ainsi un degré minimal de complication. 
Dans ce modèle, un empilement rigide de particules compress\'ees les unes contre les autres
perd sa stabilit\'e m\'ecanique lorsque la fraction d'empilement $\phi$ décroît, en dessous d'une fraction 
particulière $\phi=\phi_J$, qui dépend du protocole de préparation 
de cet empilement et de la taille du système~\cite{Chaudhuri2010}. À ce point 
particulier, la pression de confinement tend vers $0$ et les déformations des 
particules disparaissent~\cite{Liu1998,OHern2002,OHern2003,vanHecke2010}. 
Les modules élastiques se comportent en lois d'échelles avec la distance à ce point. 
Lorsque l'on approche cette transition, le matériau devient 
alors fragile~\cite{Cates1998} et sa réponse linéaire est dominée par les 
fluctuations à basses énergies~\cite{Wyart2005EPL}. \resub{Des travaux 
antérieurs sondant les propriétés mécaniques des empilements de grains 
mettaient en avant que la friction~\cite{Goldenberg2005} et les cha\^{i}nes de forces~\cite{Bouchaud2001,Kolb2006} 
sont à l'origine la nature solide de l'empilement}. Récemment, de 
nombreux efforts th\'eoriques ont montré que la nature marginale de la transition 
de blocage est l'ingrédient essentiel \`a l'origine des comportements non triviaux 
de la structure de l'empilement et de ses propriétés mécaniques et dynamiques~\cite{Charbonneau2014,Rainone2015}. 
Il s'agit de savoir si les m\^emes comportements apparaissent dans les systèmes réels. 
Plusieurs  campagnes expérimentales menées au-delà de la transition de blocage, 
ont caractéris\'e les propriétés structurelles et dynamiques 
des systèmes granulaires~\cite{Majmudar2007,Coulais2012,Coulais2014},  des mousses~\cite{Katgert2010} et des
émulsions~\cite{Jorjadze2013}. Il existe aussi quelques études qui ont détermin\'e la rhéologie
sous la transition de blocage  dans des expériences de grains vibrés~\cite{Candelier2010,Dijksman2011},
de mousses~\cite{Katgert2013}, et d'émulsions~\cite{Scheffold2013,Fall2014}. Cependant
le lien  direct avec la transition de blocage n'est pas clairement établi. En particulier, la 
pertinence de la réponse linéaire proche de la transition reste au centre des 
débats~\cite{Schreck2011,Bertrand2014,Goodrich2014}. Pour une déformation de 
cisaillement finie $\gamma$, les effets non-linéaires deviennent 
prédominants~\cite{Lerner2013,Brito2010,Gomez2012} et la réponse mécanique du système n'est 
plus uniquement décrite par la géométrie de l'empilement. Enfin, dans le cas des expériences 
avec des grains, les études numériques et théoriques de sphères molles ignorent 
systématiquement les effets de dilatance, c'est \`a dire l'accroissement de volume ou de pression sous une 
déformation de cisaillement~\cite{Reynolds1885,Bi2011,Ren2013}. \\
\indent Dans cet article, qui présente le détail d'une lettre publiée 
récemment~\cite{Coulais2014PRL}, nous présentons la première mesure expérimentale de la 
réponse élastique d'un empilement 2D de grains à travers la transition de 
blocage. Nous appliquons une déformation de cisaillement inhomogène en gonflant 
un intrus au centre d'une couche bidisperse de grains frottants. Nous 
déterminons le champ de déplacement des grains et le réseau de forces de contact 
à partir de mesures photoélastiques et des techniques de suivi de particules, et 
nous calculons le tenseur des déformations et le tenseur des contraintes à 
l'échelle du grain. Les relations constitutives obtenues à partir des courbes 
paramétriques des invariants de ces deux tenseurs montrent que l'élasticité 
linéaire ne s'applique pas. La dilatance est essentielle et, au-delà du blocage, 
une atténuation de la variation de contrainte de cisaillement se produit pour de 
faibles déformations. L'élasticité linéaire est retrouvée pour de grandes 
déformations, à partir d'une déformation critique $\gamma_c$, qui évolue avec la 
distance au blocage et qui disparaît à $\phi_J$. Nous recalculons enfin les 
profils de déformations issus des relations constitutives et nous montrons 
qu'ils correspondent aux profils expérimentaux. 

%-----------------------Setup-------------------------------------------------
\section{Dipositif expérimental et Protocole}
\begin{figure}[b!] 
\center
\includegraphics[width=\columnwidth]{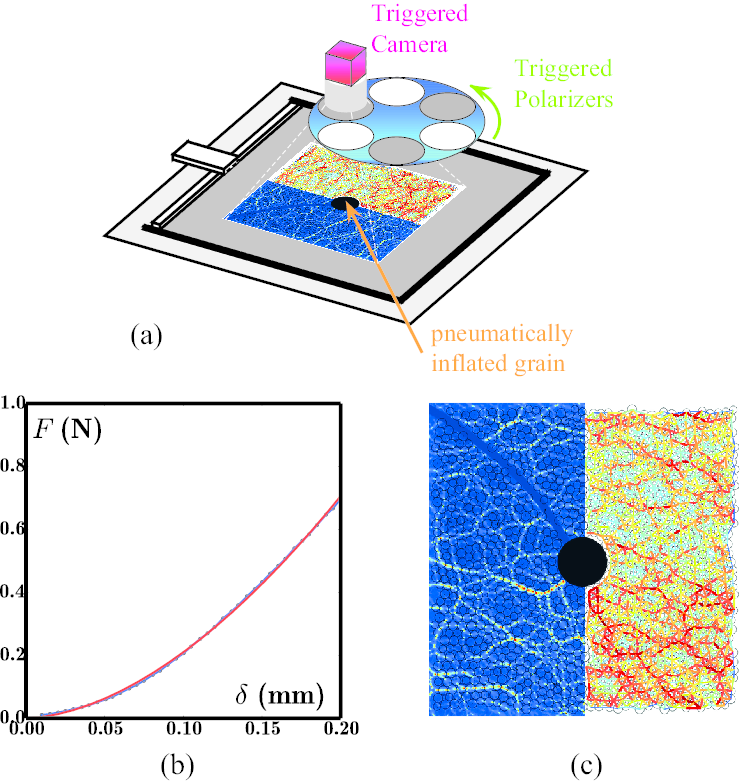}
%\vspace{-0.5cm}
\caption{{\bf Dispositif Expérimental.} (a) 8166 disques photoélastiques sont confin\'es 
dans un cadre rectangulaire, dont la surface 
est contrôlée par un piston latéral. Au centre, un "ballon" 2D est gonflé par un dispositif pneumatique, ce qui permet de créer un cisaillement
inhomogène au sein de l'assemblée. La plaque inférieure (gris clair) sur laquelle reposent les grains peut vibrer horizontalement et est utilisée 
pour homogénéiser l'assemblée entre les différents tests. (b) Test de compression uni-axiale sur un seul grain, les données brutes (ronds bleus) 
correspondent à la force $F$ (N) en fonction du déplacement $\delta$ (mm) et un ajustement (ligne rouge pleine) de la 
forme $F=A\delta^b$, avec $A=13.0\pm1.0$ et $b=1.8\pm 0.1$. (c) Photo de (gauche) données brutes entre polariseurs croisés et (droite) d'une 
reconstruction de l'assemblée de grains et du réseau de forces. La taille et la couleur des liens correspondent \`a l'intensité de la force 
normale entre chaque grain.\\
\small{\emph{{\bf Experimental Setup.} (a) 8166 photoelastic disks are confined in a rectangular frame, whose surface is tuned by a side piston.
At the center, a 2D "balloon" is inflated pneumatically, thus inducing an inhomogeneous shear in the packing.
The bottom plate (light gray), on which rest the grains, can be vibrated horizontally and is used to homogenize the packing between the different
mechanical tests. (b) Uniaxial compression test on a single grain. The raw data (blue disks) indicates the force-displacement curve.
The red solid line corresponds to a fit to $F=A\delta^b$, with $A=13.0\pm1.0$ and $b=1.8\pm 0.1$. (c) Picture of (left) the raw
cross-polarized pictures and of (right) the reconstructed grain packing and force network. The width and color of the links correspond to the
intensity of the normal force between two grains.}}
}
\vspace{-0.5cm}
\label{setup}
\end{figure}

Le dispositif est adapté de celui utilisé par~\cite{Coulais2012,Coulais2014}. Une 
couche bi-disperse de $8166$ disques photoélastiques de diamètre $4$ et $5$~mm est 
confinée dans une cellule rectangulaire. Un des murs est un piston mobile qui 
permet de contrôler précisément la fraction surfacique $\phi$. Les grains 
reposent sur une plaque en verre qui peut être vibrée à une amplitude de $1$~cm à 
une fréquence de $10$~Hz perpendiculairement à la direction du piston. L'intrus 
gonflable est une entretoise en bronze, équipée de 9 pistons radiaux,
entourée d'un joint torique de diamètre $2 r_I = 26.3$~mm et connectée au réseau 
d'air comprimé. Lorsque la pression d'air augmente dans l'intrus, les pistons 
radiaux viennent appuyer sur le joint torique, assurant une déformation radiale 
uniforme, jusqu'à $2(r_I+a) = 28.5$~mm. Quand l'alimentation en air est coupée, 
l'élasticité du joint torique permet à celui-ci de reprendre sa forme initiale. 
Le taux de dilatation de l'intrus est $a^*=a/ r_I \in [1-10]\%$. \resub{La spécificité de ce chargement 
localisé réside dans le fait qu'il sollicite le milieu granulaire selon une compression radiale 
par rapport au centre de l'intrus mais aussi selon une extension orthoradiale.}\\
\indent En variant l'amplitude de la sollicitation et la fraction volumique 
d'empilement, nous enregistrons la réponse mécanique selon un protocole précis. 
Tout d'abord, nous introduisons l'intrus au centre de l'empilement à la fraction 
d'empilement la plus basse. Ensuite, nous comprimons l'empilement à l'aide du 
piston mural jusqu'à un état fortement bloqué de l'empilement tout en vibrant la 
plaque inférieure (se référer à~\cite{Coulais2014} pour des détails 
supplémentaires). Nous arrêtons alors la vibration et commençons l'acquisition 
des images tout en augmentant la taille de l'intrus par pas de $1.5\%$ environ. 
À la fin, nous laissons l'intrus reprendre sa taille initiale, réenclenchons la 
vibration, puis décomprimons légèrement l'empilement et recommençons le cycle de 
mesure. La vibration permet de ré-homogénéiser l'état de contrainte dans 
l'empilement pendant les changements de fraction volumique, tout en maintenant 
une structure d'empilement identique~\cite{Coulais2012,Coulais2014}.\\
\indent Le processus de calibration des grains photoélastiques se fait 
indépendamment via un essai de compression uni-axiale en utilisant une machine de 
traction uni-axiale (Instron 5965) équipée d'une cellule d'effort de $100$~N. 
Nous observons que la courbe d'essai est non-linéaire, avec un exposant 
$b=1.8\pm 0.1$ (Fig.~\ref{setup}b), ce qui est cohérent avec des études 
précédentes~\cite{Huillard2011}. En revanche, ce résultat est en désaccord avec 
la théorie du contact idéal~\cite{Johnson1987}, qui prédit une loi de contact linéaire avec une 
raideur de $k=1$~N/mm pour notre système. Dans la suite, on normalisera les tenseur de contraintes
2D par cette raideur.\\
\indent Les grains photoélastiques sont éclairés par le dessous de la cellule 
avec une large source de lumière polarisée et uniforme (PHLOX). Les images sont 
enregistrées à l'aide d'une caméra CCD haute résolution ($2048\times2048$ 
pixels), conduisant à une résolution spatiale de $100\mu$m. L'acquisition de 
l'information des positions des grains et du signal photoélastique est réalisée 
à l'aide d'une roue, équipée alternativement d'un polariseur croisé, montée sur 
un moteur pas-à-pas (se référer à~\cite{Coulais2014,CoulaisPhD} pour plus de 
détails). De ces images, nous extrayons les positions des grains, que nous 
pouvons suivre pour obtenir leurs trajectoires. Ensuite, nous utilisons le champ 
de positions des grains pour construire la triangulation de Delaunay et la 
tesselation de Voronoï que nous combinons avec le signal photoélastique 
(Fig.~\ref{setup}c gauche)  afin d'estimer les forces normales et tangentielles 
entre les grains (Fig.~\ref{setup}c droite). Nous avons alors accès à un certain 
nombre d'observables à l'échelle du grain, telles que le nombre de contacts 
qu'un grain possède avec ses voisins ainsi que les tenseurs des déformations 
et des contraintes. Dans la suite, nous nous concentrons sur le nombre de contact, 
l'état de contrainte et l'état de déformation qui sont des paramètres pertinents 
pour obtenir le comportement macroscopique de l'empilement.\\

\section{\'Etat de contrainte et état de déformation à l'échelle du grain}
\begin{figure}[b!]
\centering
\includegraphics[width=\columnwidth]{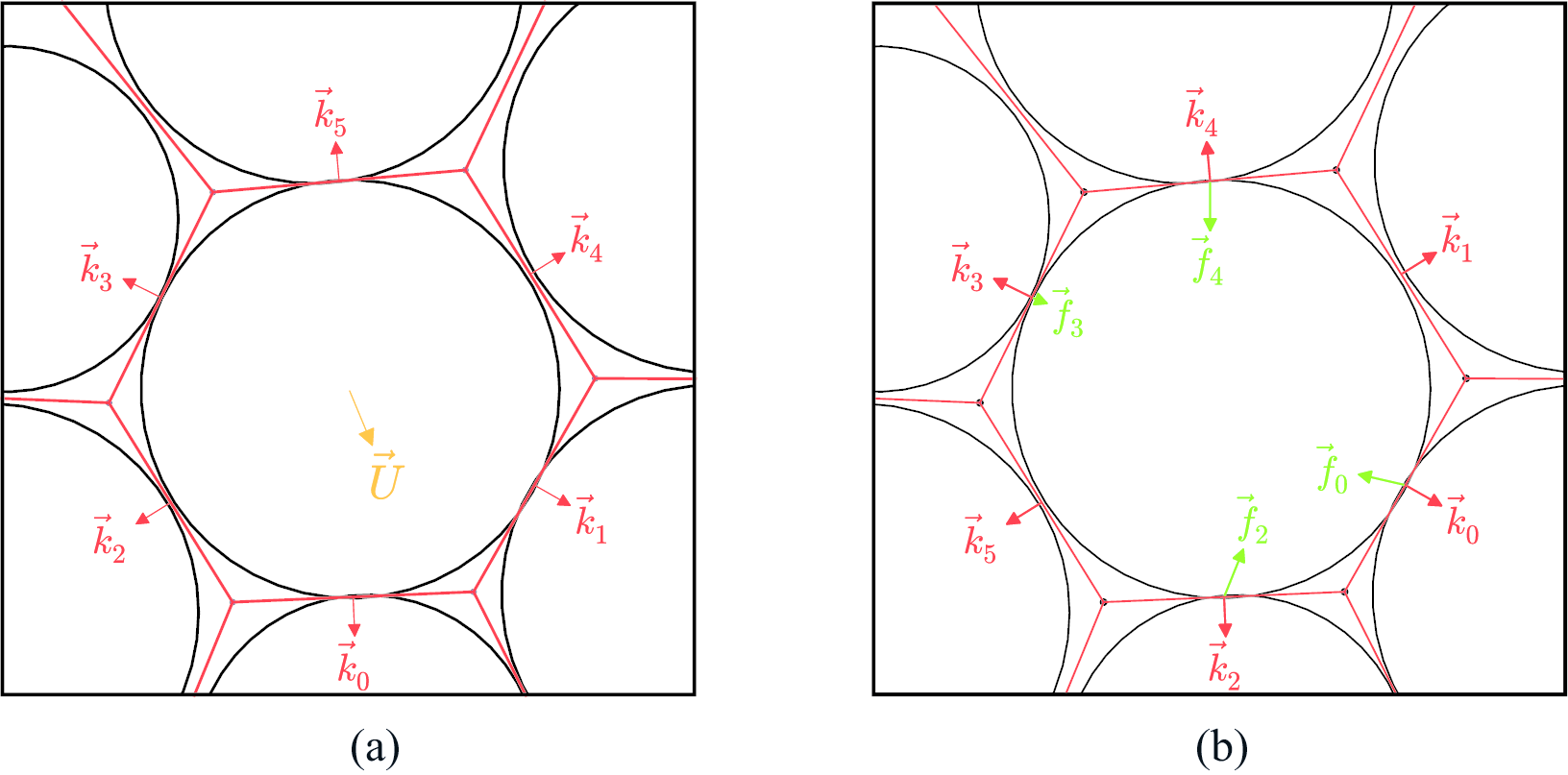}
\caption{{\bf Détermination des tenseurs de contrainte et de déformation.} (a) Le grain central 
a un  déplacement $\vec U$ et les cotés de sa cellule de Vorono\"{i} (en rouge) sont définis par 
les vecteurs $\vec k_i$. La somme du produit tensoriel de $\vec U$ avec $\vec k_i$ donne ainsi le gradient de déplacement. 
(b) Pour le tenseur des contraintes, les vecteurs $\vec k_i$ sont combinés avec les forces de contact $\vec f_i$ (en vert).\\
\small{\emph{{\bf Measurement of the stress and strain tensors.} (a) The inner grain has a displacement $\vec U$ and the
edges of its Vorono\"{i} cell (in red) are defined by the vectors $\vec k_i$.
The sum of the tensorial product between $\vec U$ and $\vec k_i$ is the displacement gradient.
(b) As for the stress tensor, the vectors $\vec k_i$ are combined with the contact forces $\vec f_i$ (in green).}}
}
\label{tensors}
\end{figure} 

\`A partir du champ de déplacement $\boldsymbol{U}$ des grains, nous calculons le 
tenseur des déformations  $\boldsymbol{\epsilon}$ à l'échelle du 
grain~\cite{Drescher1972,Cundall1982,Cambou2000}. Pour cela, nous évaluons le tenseur 
des déformations à partir du champ de déplacements des grains voisins. Cette 
formulation permet d'éviter d'effectuer un calcul par différences 
finies~\cite{Cambou2000}. Les composantes du tenseur de déformation sont estimées à 
partir du champ de déplacement $\boldsymbol{U}$ du grain $i$ et des normales aux 
cotés de sa cellule de Vorono\"{i} $S$, définies par sa position par rapport à ses voisins 
(Fig.~\ref{tensors}a). Ainsi, le gradient moyen du champ de déplacement 
de $\boldsymbol{U}$ correspondant au grain $i$ dans $S$ peut être défini par:
\begin{equation}
\langle \boldsymbol{\nabla}\boldsymbol{U} \rangle=\frac{1}{S} \iint_{S} 
\boldsymbol{\nabla}\boldsymbol{U} dS.
\end{equation}
En utilisant le théorème de la divergence, on obtient
\begin{equation}
\langle \boldsymbol{\nabla}\boldsymbol{U} \rangle
=\frac{1}{S}\oint_{L} \boldsymbol{U}  \cdot  \boldsymbol{k} dL,
\end{equation}
où $L$ est l'arête correspondant à la cellule de Voronoï et $\boldsymbol{k}$ est 
la normale sortante de cette arête (Fig.~\ref{tensors}a). 
En faisant l'hypothèse que la cellule de Voronoï est un polygone, on trouve
\begin{equation}
\langle \boldsymbol{\nabla}\boldsymbol{U} \rangle= \frac{2}{R}\sum_{L} 
\boldsymbol{U}  \cdot  \boldsymbol{k},
\end{equation}
où $R$ est le rayon du grain $i$ considéré. Le tenseur des déformations 
$\boldsymbol{\epsilon}$ correspond à la partie symétrique de ce gradient du 
champ de déplacement:
\begin{equation}
\boldsymbol{\epsilon}  = \frac{1}{2}\left(\langle \boldsymbol{\nabla} 
\boldsymbol{U} \rangle +\langle \boldsymbol{\nabla} \boldsymbol{U} 
\rangle^\dagger\right).
\end{equation}
Le principal atout de cette méthode est de ne pas utiliser de différences 
finies, ce qui réduit considérablement le bruit du calcul.\\
\indent Pour calculer le tenseur des contraintes $\boldsymbol{\sigma}$ à 
l'échelle du grain $i$, nous combinons la tesselation et les forces 
d'interactions du grains $i$ avec ses voisins selon \cite{Drescher1972,Bi2011}:
\begin{equation}
\boldsymbol{\sigma}=\frac{1}{S}\sum_{i\neq j} \boldsymbol{r}_{ij} \cdot 
\boldsymbol{f}_{ij},
\end{equation}
où $\boldsymbol{f}_{ij}$ est la force de contact entre le 
grain $i$ et un de ses voisins $j$ tandis que $\boldsymbol{r}_{ij}$ est le rayon vecteur 
orienté du centre du grain $i$ vers le centre du grain $j$ (Fig.~\ref{tensors}b).\\
\indent Après avoir vérifié que ces tenseurs partagent les mêmes vecteurs 
propres~\cite{Cortet2009}, nous restreignons l'analyse au premier et second 
invariants de chaque tenseur. Ainsi, la dilatation est définie par
 \begin{equation}
\varepsilon = \frac{1}{2} \sum\limits_{k}  \epsilon_{kk},
\end{equation}
la pression est définie par 
 \begin{equation}
P =- \frac{1}{2} \sum\limits_{k} \sigma_{kk},
\end{equation}
la déformation de cisaillement est définie par
\begin{equation}
\gamma=\sqrt{\frac{3}{2}\sum\limits_{i,j} 
\left(\epsilon_{ij}-\varepsilon\delta_{ij}\right)^2},
\end{equation}
et la contrainte de cisaillement est définie par
\begin{equation} 
\tau=\sqrt{\frac{3}{2}\sum\limits_{i,j}\left(\sigma_{ij}+P\delta_{ij}\right)^2}, 
\end{equation}
où $\delta_{ij}$ représente le symbole de Kronecker. Dans la suite, $P$ et 
$\tau$ sont normalisés par la raideur du contact $k=1$~N/mm et l'unité de 
longueur est le diamètre des petits grains $s=4$~mm. Le tenseur des contraintes 
et le tenseur des déformations sont respectivement mesurés avec une résolution 
de $10^{-4}$ et $10^{-3}$.

%-----------------------Initial State-----------------------------------------
\section{\'Etat de l'empilement non sollicité}
\begin{figure}[b!] 
\center
\includegraphics[width=\columnwidth]{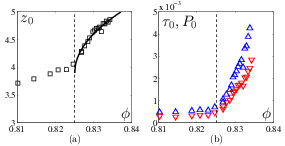}
\vspace{-0.5cm}
\caption{{\bf État de contrainte initial} (extraite de \cite{Coulais2014PRL}). \leg{(a)} 
Nombre moyen de contact initial $z_0$  ($\square$); \leg{(b)}, pression $P_0$ ($\triangle$) et contrainte 
de cisaillement $\tau_0$ ($\triangledown$) en fonction de la fraction surfacique $\phi$. 
La ligne noire est un ajustement à la fonction $z_0=z_p(\phi-\phi_J)^{0.5}+z_J$, avec
$\phi_J=0.8251\pm0.0009$, $z_p=10.0\pm 0.5$, 
et $z_J=3.9\pm 0.1$. La ligne en pointillés indique $\phi_J$.\\
\small{\emph{{\bf Initial stress state} (adapted from \cite{Coulais2014PRL}). (a) Initial average contact number
$z_0$  ($\square$); (b), pressure $P_0$ ($\triangle$) and shear stress $\tau_0$ ($\triangledown$) vs. packing
fraction $\phi$. The black line is a fit to $z_0=z_p(\phi-\phi_J)^{0.5}+z_J$, with $\phi_J=0.8251\pm0.0009$, $z_p=10.0\pm 0.5$,
and $z_J=3.9\pm 0.1$. The dash line indicates $\phi_J$.}}}
\vspace{-0.5cm}
\label{initialstress}
\end{figure}

Pour chaque fraction d'empilement, avant le gonflement de l'intrus, le système 
est caractérisé par un état initial, avec des chaînes de force se répartissant dans l'ensemble du 
matériau. La particularité de l'empilement bloqué réside dans le fait que toutes 
les forces sont répulsives, la rigidité de l'assemblée étant donnée par une 
pression de confinement. Un tel état bloqué, qui a auparavant été étudié en 
détail~\cite{Coulais2014}, est statistiquement homogène. Le nombre de contact 
moyen $z_0$ est essentiellement constant pour les faibles fractions surfaciques
(Fig.~\ref{initialstress}a). Aux fractions surfaciques moyennes, il présente une 
singularité à partir de laquelle il augmente non-linéairement. On identifie 
cette singularité avec la transition de blocage à la fraction surfacique 
$\phi_J=0.8251\pm 0.0009$. Il n'est pas surprenant d'observer une valeur non 
nulle $z_0$ sous $\phi_J$: quand la vibration est arrêtée, la structure est 
figée brutalement et des forces résiduelles demeurent à cause de la friction 
entre les grains et la paroi inférieure. La croissance non-linéaire de $z_0$ 
avec $\phi$ est compatible avec celle obtenus dans les simulations de 
particules frottantes~\cite{Katgert2010,Somfai2007} et a pour origine le 
désordre géométrique de ce type d'empilement~\cite{OHern2002,Charbonneau2014}.

 La pression initiale $P_0$ augmente aussi au-delà du blocage à partir d'une 
pression résiduelle sous $\phi_J$, qui comme pour $z_0$, est due au figeage de la structure 
(Fig.~\ref{initialstress}b). Nous comprimons l'empilement de manière non 
isotrope, en déplaçant uniquement le piston mural. Malgré nos efforts pour 
préparer le matériau avec autant de précaution que faire se peut~\cite{Coulais2012,Coulais2014}, 
en utilisant un protocole de compactification logarithmique sur une journée, 
l'empilement conserve de l'anisotropie clairement mis en évidence 
par l'existence d'une contrainte de cisaillement résiduelle $\tau_0$ 
proportionnelle à la pression $P_0$ (Fig.~\ref{initialstress}b). Cependant, le 
rapport $\tau_0/P_0 < 1$, ce qui est attendu pour les 
empilements où la contrainte de compression domine. Dans notre système, la 
sollicitation localisée est axisymétrique. Par conséquent, de telles 
fluctuations anisotropes~\cite{Goodrich2014PRE} 
sont moyennées et notre mesure est robuste face \`a l'anisotropie. Il est \`a noter que
la plupart des empilements préparés numériquement présentent aussi 
des contraintes de cisaillement résiduelles et une forte anisotropie~\cite{DagoisBohy2012,Goodrich2014PRE}.

%-----------------------Response to inflation----------------------------------
\section{Réponse au gonflement}

\begin{figure}[b!] 
\center
\includegraphics[width=\columnwidth]{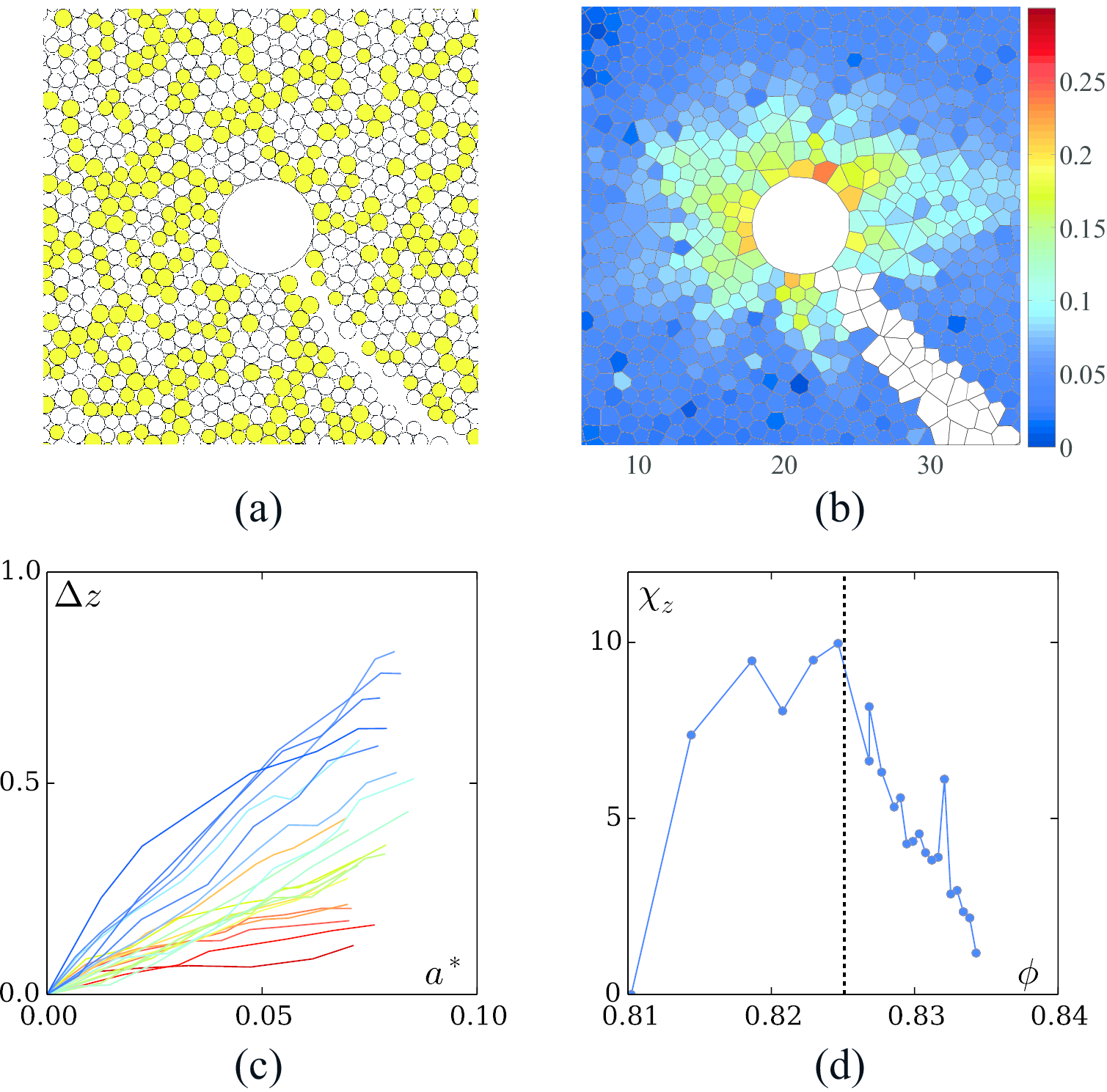}
\caption{{\bf Augmentation du nombre de contacts.}
Cartographies (a) des changements du nombre de contacts et (b) des déplacements radiaux $u_r$ pour 
$\phi=0.8294$ et $a^*=4.4\%$. Chaque grain coloré en jaune a gagné ou perdu au moins un contact.  Les grains sans couleurs sont masqués 
par le tube pneumatique qui alimente l'intrus en air comprimé.
\leg{(c):}  Différence du nombre de contact moyen avec l'état initial, $\Delta z$ en fonction du paramètre
de gonflement $a^*=\frac{a}{R_0}$ pour différentes fractions surfaciques. Le code 
couleurs s'échelonne de bleu pour les faibles fractions surfaciques à rouge pour les grandes 
fractions surfaciques. \leg{(d):} Pente $\chi_z$ de la courbe  $\Delta z$-$a^*$ en fonction de la fraction 
surfacique, $\phi$.\\
\small{\emph{{\bf Increase of contact number.}
Maps (a) of contact changes and (b) of radial displacements $u_r$ for $\phi=0.8294$ and
$a^*=4.4\%$. The yellow grains have lost or gained at least one contact. The uncolored grains sit below
the pneumatic tube connected to the intruder, which masks the field of view.
(c) Difference between the average contact number with the initial state, $\Delta z$ vs. the inflation parameter
$a^*=\frac{a}{R_0}$ for various packing fractions. Color code spans from blue to red with increasing
packing fractions. (d) Slope $\chi_z$ of the $\Delta z$-$a^*$ curve vs. the packing fraction, $\phi$.}}
}
\label{chiz}
%\vspace{-0.8cm}
\end{figure}

Dans un premier temps, nous allons étudier le réseau de contact lors du 
gonflement de l'intrus. La figure~\ref{chiz} présente le changement de contact dans 
ce réseau pendant la phase de compression. On remarque que ce changement de 
contact concerne plusieurs grains, répartis de manière homogène dans 
l'empilement (fig.~\ref{chiz}a) . \resub{Ceci est d'autant plus remarquable que la sollicitation 
elle-même n'engendre pas une déformation homogène de l'empilement: le déplacement des grains est d'autant plus faible que l'on s'éloigne du centre de l'intrus (fig.~\ref{chiz}b)}.

Ceci suggère fortement que la réponse mécanique est dominée par le désordre. 
Le nombre de changement de contact moyen par grain $\Delta z$ croît linéairement 
avec le paramètre de dilatation $a^*$ (Fig.~\ref{chiz}c), avec une pente 
positive qui dépend de $\phi$. Donc, lors du gonflement de l'intrus et du 
cisaillement de la structure, des contacts sont en moyenne créés, ce qui est en accord avec 
le fait que la surface totale occupée par les grains diminue, mais sans aucune structure 
spatiale ni corrélation avec le forçage mécanique.
Nous modélisons cet accroissement par une fonction linéaire et nous reportons la pente 
$\chi_z$ dans la figure~\ref{chiz}d en fonction de $\phi$. Premièrement, 
$\chi_z$ est toujours strictement positif. Cela suggère que de l'énergie 
pourrait être dissipée lors d'un tel processus et que des effets non-linéaires 
sont attendus. Ensuite, la courbe présente un maximum précisément à $\phi_J$, 
qui est la fraction volumique de blocage. $\phi_J$ joue donc un rôle précis dans la 
réponse mécanique de l'empilement. De telles observations pourraient fournir de 
précieux indices pour la formulation d'un modèle micro-mécanique.\\
\indent Dans la suite, nous allons considérer une formulation plus classique des tenseurs 
de contraintes et de déformations en utilisant leurs invariants. Nous allons 
considérer l'excès de pression $P$ et de contrainte de cisaillement $\tau$ par 
rapport à l'état de contrainte initial pour l'ensemble des $a^*$ et $\phi$ 
mesurés. Le comportement d'un matériau élastique linéaire, homogène et isotrope 
classique a une relation linéaire entre l'état de contrainte et l'état 
de déformation, ce qui peut s'exprimer grâce aux relations $P =- K  \varepsilon$ et 
$\tau = 2G\gamma$, où $K$ et $G$ sont respectivement le module de 
compressibilité et le module de cisaillement. Dans l'hypothèse d'une géométrie 
2D axisymétrique, l'intégration de l'équation d'équilibre conduit à  $ \tau \sim 
G \gamma \sim a^* /r^2$, où $r$ est la variable d'espace: la 
contrainte de cisaillement diminue avec la distance au centre de l'intrus. En revanche, 
la pression $P \sim 1$ ne dépend pas de l'espace. \resub{Ainsi, dans cette formulation, le forçage localisé induit des déformations 
volumiques et cisaillantes.}

\begin{figure}[b!] 
\center
\includegraphics[width=\columnwidth]{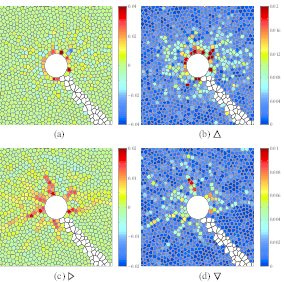}
\vspace{-0.5cm}
\caption{{\bf Cartographies des invariants des déformations et des contraintes} (extraite de \cite{Coulais2014PRL}). 
Cartographies de la dilatation, $\varepsilon$ ,\leg{(a)}, cisaillement, $\gamma$, \leg{(b)}, 
pression, $P$, \leg{(c)} et contrainte de cisaillement, $\tau$, \leg{(d)}, pour 
$\phi=0.8294$ et $a^*=4.4\times10^{-2}$. Les images représentées ici ne représentent 
qu'un tiers du système entier.\\
\small{\emph{{\bf Maps of the strain and stress invariants.} (Adapted from \cite{Coulais2014PRL}). Maps of
dilation, $\varepsilon$ ,\leg{(a)}, shear strain, $\gamma$, \leg{(b)}, pressure,
$P$, \leg{(c)} and shear stress, $\tau$, \leg{(d)}, for $\phi=0.8294$ and
$a^*=4.4\times10^{-2}$. The uncolored grains sit below the pneumatic
tube connected to the intruder, which masks the field of view.}}
}
\vspace{-0.5cm}
\label{Maps}
\end{figure}

Nous montrons en figure~\ref{Maps} les quatre cartes typiques des invariants 
pour une fraction d'empilement $\phi=0.8294$ supérieure à $\phi_J$ et un taux de 
dilatation $a^*$ $(4.4\times 10^{-2})$. Nous observons de fortes fluctuations 
spatiales et la réponse s'écarte de la réponse linaire élastique précisée plus 
tôt. Cet écart est inhérent aux inhomogénéités présentes dans les matériaux 
désordonnés. Cependant, en considérant des échelles de longueurs plus grandes, 
on observe que l'axisymétrie du chargement est globalement respectée et que 
l'intensité de la réponse décroit avec la distance au centre de l'intrus. On 
peut ainsi faire l'hypothèse que notre cellule est suffisamment grande pour ne pas sentir les effets de confinement. 
\resub{Proche de l'intrus, une importante dilatation se produit à cause
de la condition limite impos\'ee par l'intrus lui-même. Etant donné que l'intrus possède un diamètre plus important que celui des grains, la variation de fraction surfacique localisée due au gonflement reste singulière par rapport à la fraction surfacique moyenne.}  Le reste de l'empilement présente de large fluctuations (avec ici un écart type 
de $3\times 10^{-3}$) avec un léger accroissement de la fraction surfacique 
(d'amplitude $6\times 10^{-5}$)  (Fig.~\ref{Maps}a), assurant la conservation 
globale du volume.

\begin{figure}[b!] 
\center
\includegraphics[width=\columnwidth]{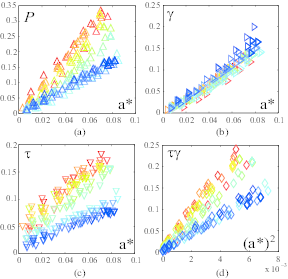}
\vspace{-0.5cm}
\caption{{\bf R\'eponse au gonflement.} Moyennes spatiales des invariants des tenseurs 
en fonction du paramètre de gonflement $a^*$. (a) Pression $P$, (b) 
Cisaillement $\gamma$, (c) Contrainte de cisaillement $\sigma$ et (d) puissance de travail des efforts de cisaillement 
$\gamma\sigma$. Le code couleurs s'échelonne de bleu pour les faibles fractions surfaciques à rouge 
pour les grandes fractions surfaciques.\\
\small{\emph{{\bf Response to inflation.} Spatial averaged of the tensors vs. the inflation parameter $a^*$. (a) Pressure $P$, (b)
shear strain $\gamma$, (c) shear stress $\sigma$ and (d) shear stress power $\gamma\sigma$.
Color code spans from blue to red with increasing packing fractions.}}}
\vspace{-0.5cm}
\label{energy}
\end{figure}

Par conséquent, en dehors de la première couronne de grains autour de l'intrus, 
que nous allons exclure de l'analyse, le matériau peut être considéré comme 
incompressible. Nous postulerons donc dans la suite que $\varepsilon = 0$. \resub{Ainsi, la sollicitation mécanique se révèle plutôt être une déformation cisaillante pure.}
La 
deuxième observation importante montre un écart significatif \`a la réponse élastique linéaire: 
le champ de pression est inhomogène et diminue à mesure que 
l'on s'écarte du centre de l'intrus. Comme ce champ de pression ne correspond pas à des variations 
sensibles de volume, il est n\'ecessairement induit par la déformation de cisaillement. Cet 
effet est la manifestation de la pr\'esence de dilatance dans un système à volume constant, un effet 
bien connu dans les milieux granulaires~\cite{Reynolds1885}. Le coefficient de dilatance à pression 
constante est défini par $P=R\gamma^2$, et lié à celui défini à volume constant 
$D$ par le module de compressibilité $K$ tel que $R=DK$~\cite{Tighe2014}.\\
\indent Finalement, tandis que la pression $P$ (moyennée ortho-radialement) 
évolue linéairement avec $a^*$ (Fig.~\ref{energy}a), la déformation de 
cisaillement $\gamma$ (moyennée ortho-radialement) croît plus vite avec $a^*$ 
(Fig.~\ref{energy}b). C'est une indication de la nature non-linéaire du 
matériau. En revanche, le travail des efforts de cisaillement $\tau\gamma$ évolue en ${a^*}^2$ 
(Fig.~\ref{energy}d), ce qui suggère que la réponse mécanique du milieu granulaire 
à la sollicitation est dominée par des déformations élastiques. Ainsi, malgré le changement 
significatif de contacts dans l'empilement, il n'y a essentiellement pas de dissipation de l'\'energie. 
Ces observations sont valables pour l'ensemble des fractions surfaciques considérées. 

%-----------------------Constitutive laws------------------------------
\section{Relations constitutives} 
\begin{figure}[t!] 
\center
\includegraphics[width=\columnwidth]{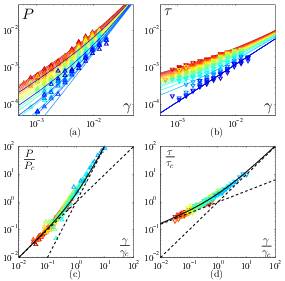}
\vspace{-0.5cm}
\caption{{\bf Relations de comportement} (figure adaptée de \cite{Coulais2014PRL}). Pression, $P$ \leg{(a)}, et 
contrainte de cisaillement, $\tau$ \leg{(b)},  en fonction du cisaillement $\gamma$, pour 21 fractions surfaciques 
$\phi\in [0.8102-0.8343]$. Les lignes pleines sont données par les équations~(\ref{eq:law1}-\ref{eq:law2}). 
Le code couleurs s'échelonne de bleu pour les faibles fractions surfaciques à rouge pour les grandes 
fractions surfaciques.
\leg{(c)} et \leg{(d)}: mêmes données que \leg{(a)} et \leg{(b)} normalisées par $\gamma_c(\phi), P_c(\phi)$ et 
$\tau_c(\phi)$. 
Les lignes pleines sont données par la version normalisée des équations.~(\ref{eq:law1}-\ref{eq:law2}) 
et les lignes en pointillés indiquent les régimes asymptotiques.\\
\small{\emph{{\bf Constitutive laws.} (Adapted from \cite{Coulais2014PRL}). Pressure, $P$ \leg{(a)}, and shear stress,
$\tau$ \leg{(b)},  vs. shear strain, $\gamma$, for 21 packing fractions $\phi
\in [0.8102-0.8343]$. The solid lines are given by Eqs.~(\ref{eq:law1}-\ref{eq:law2}).
Color code spans from blue to red with increasing packing fractions.
\leg{(c)} and \leg{(d)}: same data as \leg{(a)} and \leg{(b)} rescaled by $\gamma_c(\phi), P_c(\phi)$ and
$\tau_c(\phi)$. The solid lines are given by the rescaled version of Eqs.~(\ref{eq:law1}-\ref{eq:law2}) and the dashed lines indicate  the asymptotic regimes.
}}}
\vspace{-0.5cm}
\label{Jamming_cross}
\end{figure}

Nous détaillons \`a pr\'esent l'analyse quantitative des relations de comportement entre 
les invariants $\tau$, $P$ et $\gamma$. Nous rassemblons l'ensemble des données 
moyennées orthoradialement $P(r,a^*)$ et $\tau(r,a^*)$ en fonction de 
$\gamma(r,a^*)$, où $r$ est la distance au centre de l'intrus. 
Les figures~\ref{Jamming_cross}a et b montrent la contrainte de cisaillement et la 
pression en fonction de la déformation de cisaillement $\gamma$ pour différentes 
fractions surfaciques $\phi$. Sous $\phi_J$, la pression $P$ et la contrainte de 
cisaillement $\tau$ présentent la dépendance attendue avec la déformation de 
cisaillement $\gamma$: $\tau = 2 G_0 \gamma$ et $P=R_0 \gamma^2$, où $G_0$ et 
$R_0$ ne dépendent pas de $\phi$. Au dessus de $\phi_J$, des non-linéarités 
apparaissent sous la forme d'une diminution de la variation de $\tau$ et $P$ 
avec $\gamma$. La transition de blocage joue un rôle essentiel et nous l'utilisons 
pour décrire nos données avec un unique jeu de paramètres définis par
\ba
 P & = & \left[R_0+R_{nl}(\Delta\phi,\gamma)\right]\, \gamma^2 \label{eq:law1}\\
 \tau & = &2 \left[G_0+G_{nl}(\Delta\phi,\gamma)\right]\, \gamma \label{eq:law2}
\ea  
\noindent avec $\Delta\phi = \phi - \phi_J$, $G_0=6.0\pm0.2\times10^{-2}$, $R_0=1.2\pm0.1\times 10^1$ et
\ba
R_{nl}(\Delta\phi,\gamma) & = & \left\lbrace\begin{array}{c}
                              0 \quad\quad\quad\quad\quad  {\rm for}\, \phi < \phi_J \nonumber \\
                              a \Delta\phi^\mu \gamma^{\alpha-2} \quad {\rm for}\, \phi > \phi_J
                            \end{array} \right. ,
                            \\
G_{nl}(\Delta\phi,\gamma) & = & \left\lbrace\begin{array}{c}
			       0 \quad\quad\quad\quad\quad {\rm for}\, \phi < \phi_J \nonumber \\
			        b \Delta\phi^\nu \gamma^{\beta-1}\quad  {\rm for}\, \phi > \phi_J
                                \end{array} \right. ,
\ea
\noindent

\noindent
avec $\mu=1.7\pm 0.1,\, \alpha=1.0\pm 0.1,\, a= 8.1\pm 0.3\times10^{-2},\, 
\nu=1.0\pm 0.1,\, \beta=0.4\pm 0.1,\,
b=7.5\pm 0.3\times10^{-1}$. \`A partir de ces relations, on peut tracer deux 
courbes maîtresses présentées dans les figures~\ref{Jamming_cross}c et d avec 
$\gamma_c \sim \Delta\phi^\zeta$, $\tau_c=2G_0 \gamma_c$ et $P_c=R_0 
\gamma_c^2$. Bien que les exposants $(\mu,\alpha)$ et $(\nu,\beta)$ aient été 
obtenus indépendamment, il est remarquable que $\zeta = \mu/(2-\alpha)$ et 
$\zeta =\nu/(1-\beta)$ conduisent à la même valeur $\zeta = 1.7$. Par 
conséquent, les données de $P$ et $\tau$ présentent un comportement non-linéaire 
similaire et peuvent être décrites de la même manière. Les équations 
(\ref{eq:law1}-\ref{eq:law2}) et les lois d'échelles associées sont des 
résultats clés de l'étude. Il est à noter que le régime linéaire observé ici ne 
doit pas être confondu avec la réponse linéaire puisqu'elle se produit pour de 
grandes déformations. Nous expliquons la présence de ce nouveau régime par une 
saturation des non-linéarités. Nous pensons qu'un tel régime, qui existe aussi 
pour les empilements non bloqués, mérite une étude approfondie. Pour de petites 
déformations $(\gamma \simeq 10^{-6})$, sondées dans les études 
numériques~\cite{OHern2003,DagoisBohy2012} mais plus faibles que les plus 
petites déformations sondées dans l'expérience, on s'attend à retrouver la 
véritable réponse linéaire pour tout $\Delta\phi>0$~\cite{Goodrich2014}. Pour 
des déformations du même ordre de grandeur que celles de l'expérience, des études 
numériques récentes ont rapporté l'existence d'un changement de régime: pour des déformations plus grandes, 
la réponse devient non-linéaire, avec un exposant $\beta\simeq 0.5$~\cite{Otsuki2014,Tighe2014}, 
ce qui est compatible avec le régime non-linéaire observé ici.
 Dans de telles études, des particules avec un potentiel de déformation harmonique sont utilisées, {\em i.e.} 
$P_0=\Delta \phi$ et le croisement $\gamma^*=P_0=\Delta \phi^b$, où $b=1$. Ici, 
nos grains ne sont pas harmoniques et leur contact a un exposant $b=1.8$. En outre, nous 
observons $\gamma_c\sim \Delta \phi^\zeta$, avec $\zeta\approx b$, ce qui induit 
$\gamma_c\sim \gamma^*$. Ces deux seuils sont différents, mais ils ont le même exposant, ce qui
suggère qu'il se comportent de la même manière vis-à-vis de la 
transition de blocage. Cette étude préconise des investigations plus poussées, 
en particulier dans le régime non-linéaire, pour dévoiler le rôle des non-linéarités 
dans les lois de contact ainsi que celui de la friction entre grains. 
De plus, nous observons un rôle important de la dilatance dans nos grains 
cisaillés, qui montrent que de tels effets non-linéaires doivent être pris en 
compte pour décrire complètement le matériau. L'importance de la dilatance dans 
les solides marginaux a récemment été mise en lumière dans~\cite{Tighe_granularmatter2014}, où il a 
été montré que le coefficient de Reynolds à volume constant $R_V$ se comporte 
comme $\Delta\phi^{-1/2}$. Ici, nous reportons aussi un comportement singulier, mais 
qui concerne les propriétés non-linéaires et en particulier l'adoucissement de la 
dilatance. Dans un contexte un peu différent, Ren et al.~\cite{Ren2013} ont observ\'e 
une très forte augmentation de la dilatance sous cisaillement homogène lorsque la transition de blocage est 
approchée par le dessous. La valeur du coefficient de dilatance mesuré ici est très grande ($R_0\sim 10^4$~N/m), 
et pourrait correspondre à une saturation de la divergence observée par Ren~\cite{Ren2013}.

%----------------------- Self-consistency check---------------------------------
\section{Profils de déformation}
Nous procédons à une vérification de l'auto-cohérence du modèle en injectant 
les relations constitutives du modèle pour calculer les profils $\gamma$ et les 
comparer aux mesures expérimentales. La géométrie axisymétrique assure que la 
distance au centre $r$ de l'intrus est le seul paramètre d'espace dont dépend 
$\boldsymbol{\epsilon}$ et  $\boldsymbol{\sigma}$. Ainsi, en coordonnées 
cylindriques, les relations constitutives peuvent s'écrire sous la forme 
tensorielle

\begin{equation}
\begin{split}
\boldsymbol{\sigma}=-P_c\left(\left(\frac{\gamma_c}{\gamma}\right)^\alpha+\left(\frac{\gamma_c}{\gamma}\right)^2\right)\left[\begin{array}{cc}                                                                             1&0\\
0&1                                                                              \end{array}\right]\\
+\tau_c \left(\left(\frac{\gamma}{\gamma_c}\right)^{\beta}+\frac{\gamma}{\gamma_c}\right)\left[\begin{array}{cc}                                                                             -1&0\\
0&1                                                                              \end{array}\right].
\end{split}
\nonumber
\end{equation}
Afin de simplifier, nous choisissons d'introduire la déformation de cisaillement réduite $\tilde{\gamma}=\gamma/\gamma_c$. La relation précédente devient
\begin{equation}
\boldsymbol{\sigma}=-P_c\left(\tilde{\gamma}^\alpha+\tilde{\gamma}^2\right)\left[\begin{array}{cc}                                                                             1&0\\
0&1                                                                              \end{array}\right]+\tau_c \left(\tilde{\gamma}^{\beta}+\tilde{\gamma}\right)\left[\begin{array}{cc}                                                                             -1&0\\
0&1                                                                              \end{array}\right].
\nonumber
\end{equation}

\begin{figure}[t!] 
\center
\includegraphics[width=\columnwidth]{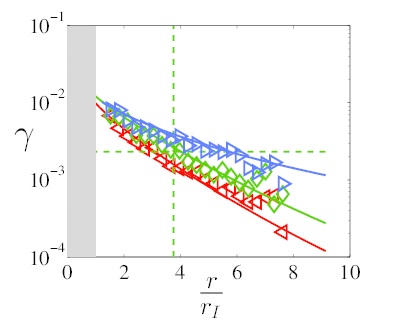}
\caption{{\bf Profils de cisaillement} (extraite de \cite{Coulais2014PRL}).
Profils de cisaillement pour 
($\triangleright$) ($\phi=0.8208; a^*=0.0374$), ($\diamond$)
($\phi=0.8268;a^*=0.0314$) et ($\triangleleft$) ($\phi=0.8338;a^*=0.0306$). 
Les marqueurs correspondent aux données expérimentales et les lignes pleines à l'intégration numérique de 
l'équation~(\ref{profile}). 
La ligne verte pointillée indique le seuil $(r_c,\gamma_c)$ pour $\phi=0.8268$ et $a^*=0.0314$.
La zone grisée correspond a la zone occupée par le "ballon".\\
\small{\emph{{\bf Shear strain profiles} (Adapted from \cite{Coulais2014PRL}).
Shear strain profile for ($\triangleright$) ($\phi=0.8208; a^*=0.0374$), ($\diamond$)
($\phi=0.8268;a^*=0.0314$) and ($\triangleleft$) ($\phi=0.8338;a^*=0.0306$).
The symbols are experimental data and the solid lines come from the integration of eq.(\ref{profile}).
The green dashed line indicates the crossover for the case ($\phi=0.8268;a^*=0.0314$).}}}
\vspace{-0.5cm}
\label{summary}
\end{figure}
\noindent
L'équilibre mécanique se traduit par $\boldsymbol{\nabla}\boldsymbol{\cdot}\boldsymbol{\sigma} =\boldsymbol{0}$, et conduit à
\begin{equation}
-P_c\left(\alpha\tilde{\gamma}^{\alpha-1}+2\tilde{\gamma}\right)\frac{d\tilde{\gamma}}{dr}-\tau_c\left(\beta\tilde{\gamma}^{\beta-1}+1\right)\frac{d\tilde{\gamma}}{dr}-2\tau_c\frac{\tilde{\gamma}^{\beta}+\tilde{\gamma}}{r}=0,
\nonumber
\end{equation}
qui, par séparation des variables $\tilde{\gamma}$ et $r$, peut être réécrit en 
\begin{equation}
\frac{P_c (\alpha\tilde{\gamma}^{\alpha-1} + 2\tilde\gamma) + \tau_c
(\beta\tilde{\gamma}^{\beta-1}+1)}{\tilde{\gamma}^{\beta}+\tilde{\gamma}} 
d\tilde{\gamma} = - 2\tau_c\frac{dr}{r}.
\label{profile}
\end{equation}
Nous intégrons numériquement l'équation~\ref{profile} avec la condition limite
$\tilde\gamma(r=r_I)=a^*/\gamma_c$ et nous obtenons les profils tracés sur la 
figure~\ref{summary}, à laquelle nous avons ajouté les données expérimentales. 
L'accord est satisfaisant, étant donné qu'il n'y a aucun paramètre ajustable et 
que nous avons négligé le confinement à grand $r$. Ceci confirme que notre 
description des relations constitutives est cohérente. 

%-----------------------Discussion ---------------------------------------------
\vspace{-0.5cm} 
\section{Conclusion}
Dans ce travail, nous avons fourni une caractérisation quantitative de la réponse 
élastique d'un empilement 2D de grains proche de la transition de blocage au 
gonflement local d'un intrus. La géométrie particulière sonde la réponse à un 
cisaillement inhomogène à volume constant \resub{et peut \^{e}tre vue comme un analogue \`a sym\'etrie cylindrique du 
probl\`eme de l'inclusion d'Eshelby. Cette \'etude pourrait ainsi trouver une analogie avec les probl\`emes 
de plasticit\'e o\`u les contraintes \'elastiques sont redistribu\'ees par un \'ev\'enement plastique localis\'e~\cite{Talamali2008,tsamados2009local}.} 

Nos résultats mettent en lumière 
l'effet de dilatance et dévoilent un régime non-linéaire au delà de la transition 
de blocage où le module de cisaillement et le module de dilatance diminuent 
jusqu'à ce qu'un nouveau régime linéaire apparaisse pour de grandes 
déformations.
Cette étude montre l'existence d'un 
cisaillement critique $\gamma_c$ qui sépare le régime non linéaire d'un régime 
linéaire saturé. Son comportement avec la distance au blocage est cohérent avec 
un autre croisement $\gamma^*$, récemment identifié lorsque l'élasticité passe du 
régime linéaire au régime non-linéaire~\cite{Tighe2014}, mais est tel que  $\gamma^*\ll\gamma_c$.
Enfin, notre étude fournit les fondations pour un modèle continu de tels matériaux. En 
particulier, l'observation du changement de réseau de contact est une 
observation clé pour pouvoir à terme obtenir une description micro-mécanique avec 
un formalisme de Cosserat~\cite{Cosserat1909}.

%-----------------------Acknowledgements----------------------------------------

{\em Remerciements} --- Nous remercions B. Tighe, W. Ellenbroek et M.
van Hecke pour des discussions importantes. Nous sommes très reconnaissants à V. 
Padilla et C. Wiertel-Gasquet pour l'assistance et le support technique. Ce 
travail est financé par le projet ANR STABINGRAM No. 2010-BLAN-0927-01 ainsi que 
par les projets REMIGS2D and COMIGS2D du RTRA Triangle de la Physique.

%-----------------------Bibliographie----------------------------------------
\bibliography{bibliomattech}

\begin{thebibliography}{48}%
\makeatletter
\providecommand \@ifxundefined [1]{%
 \@ifx{#1\undefined}
}%
\providecommand \@ifnum [1]{%
 \ifnum #1\expandafter \@firstoftwo
 \else \expandafter \@secondoftwo
 \fi
}%
\providecommand \@ifx [1]{%
 \ifx #1\expandafter \@firstoftwo
 \else \expandafter \@secondoftwo
 \fi
}%
\providecommand \natexlab [1]{#1}%
\providecommand \enquote  [1]{``#1''}%
\providecommand \bibnamefont  [1]{#1}%
\providecommand \bibfnamefont [1]{#1}%
\providecommand \citenamefont [1]{#1}%
\providecommand \href@noop [0]{\@secondoftwo}%
\providecommand \href [0]{\begingroup \@sanitize@url \@href}%
\providecommand \@href[1]{\@@startlink{#1}\@@href}%
\providecommand \@@href[1]{\endgroup#1\@@endlink}%
\providecommand \@sanitize@url [0]{\catcode `\\12\catcode `\$12\catcode
  `\&12\catcode `\#12\catcode `\^12\catcode `\_12\catcode `\%12\relax}%
\providecommand \@@startlink[1]{}%
\providecommand \@@endlink[0]{}%
\providecommand \url  [0]{\begingroup\@sanitize@url \@url }%
\providecommand \@url [1]{\endgroup\@href {#1}{\urlprefix }}%
\providecommand \urlprefix  [0]{URL }%
\providecommand \Eprint [0]{\href }%
\providecommand \doibase [0]{http://dx.doi.org/}%
\providecommand \selectlanguage [0]{\@gobble}%
\providecommand \bibinfo  [0]{\@secondoftwo}%
\providecommand \bibfield  [0]{\@secondoftwo}%
\providecommand \translation [1]{[#1]}%
\providecommand \BibitemOpen [0]{}%
\providecommand \bibitemStop [0]{}%
\providecommand \bibitemNoStop [0]{.\EOS\space}%
\providecommand \EOS [0]{\spacefactor3000\relax}%
\providecommand \BibitemShut  [1]{\csname bibitem#1\endcsname}%
\let\auto@bib@innerbib\@empty
%</preamble>
\bibitem [{\citenamefont {Chaudhuri}\ \emph {et~al.}(2010)\citenamefont
  {Chaudhuri}, \citenamefont {Berthier},\ and\ \citenamefont
  {Sastry}}]{Chaudhuri2010}%
  \BibitemOpen
  \bibfield  {author} {\bibinfo {author} {\bibfnamefont {P.}~\bibnamefont
  {Chaudhuri}}, \bibinfo {author} {\bibfnamefont {L.}~\bibnamefont {Berthier}},
  \ and\ \bibinfo {author} {\bibfnamefont {S.}~\bibnamefont {Sastry}},\
  }\bibfield  {title} {\enquote {\bibinfo {title} {Jamming transitions in
  amorphous packings of frictionless spheres occur over a continuous range of
  volume fractions},}\ }\href {\doibase 10.1103/PhysRevLett.104.165701}
  {\bibfield  {journal} {\bibinfo  {journal} {Phys. Rev. Lett.}\ }\textbf
  {\bibinfo {volume} {104}},\ \bibinfo {pages} {165701} (\bibinfo {year}
  {2010})}\BibitemShut {NoStop}%
\bibitem [{\citenamefont {Liu}\ and\ \citenamefont {Nagel}(1998)}]{Liu1998}%
  \BibitemOpen
  \bibfield  {author} {\bibinfo {author} {\bibfnamefont {A.~J.}\ \bibnamefont
  {Liu}}\ and\ \bibinfo {author} {\bibfnamefont {S.~R.}\ \bibnamefont
  {Nagel}},\ }\bibfield  {title} {\enquote {\bibinfo {title} {Nonlinear
  dynamics: Jamming is not just cool any more},}\ }\href {\doibase
  10.1038/23819} {\bibfield  {journal} {\bibinfo  {journal} {Nature}\ }\textbf
  {\bibinfo {volume} {396}},\ \bibinfo {pages} {21--22} (\bibinfo {year}
  {1998})}\BibitemShut {NoStop}%
\bibitem [{\citenamefont {O'Hern}\ \emph {et~al.}(2002)\citenamefont {O'Hern},
  \citenamefont {Langer}, \citenamefont {Liu},\ and\ \citenamefont
  {Nagel.}}]{OHern2002}%
  \BibitemOpen
  \bibfield  {author} {\bibinfo {author} {\bibfnamefont {C.~S.}\ \bibnamefont
  {O'Hern}}, \bibinfo {author} {\bibfnamefont {S.~A.}\ \bibnamefont {Langer}},
  \bibinfo {author} {\bibfnamefont {A.~J.}\ \bibnamefont {Liu}}, \ and\
  \bibinfo {author} {\bibfnamefont {S.~R}\ \bibnamefont {Nagel.}},\ }\bibfield
  {title} {\enquote {\bibinfo {title} {Random packings of frictionless
  particles},}\ }\href {\doibase 10.1103/PhysRevLett.88.075507} {\bibfield
  {journal} {\bibinfo  {journal} {Phys. Rev. Lett.}\ }\textbf {\bibinfo
  {volume} {88}},\ \bibinfo {pages} {075507} (\bibinfo {year}
  {2002})}\BibitemShut {NoStop}%
\bibitem [{\citenamefont {O'Hern}\ \emph {et~al.}(2003)\citenamefont {O'Hern},
  \citenamefont {Silbert}, \citenamefont {Liu},\ and\ \citenamefont
  {Nagel}}]{OHern2003}%
  \BibitemOpen
  \bibfield  {author} {\bibinfo {author} {\bibfnamefont {C.~S.}\ \bibnamefont
  {O'Hern}}, \bibinfo {author} {\bibfnamefont {L.~E.}\ \bibnamefont {Silbert}},
  \bibinfo {author} {\bibfnamefont {A.~J.}\ \bibnamefont {Liu}}, \ and\
  \bibinfo {author} {\bibfnamefont {S.~R.}\ \bibnamefont {Nagel}},\ }\bibfield
  {title} {\enquote {\bibinfo {title} {Jamming at zero temperature and zero
  applied stress: The epitome of disorder},}\ }\href {\doibase
  10.1103/PhysRevE.68.011306} {\bibfield  {journal} {\bibinfo  {journal} {Phys.
  Rev. E}\ }\textbf {\bibinfo {volume} {68}},\ \bibinfo {pages} {011306}
  (\bibinfo {year} {2003})}\BibitemShut {NoStop}%
\bibitem [{\citenamefont {van Hecke}(2010)}]{vanHecke2010}%
  \BibitemOpen
  \bibfield  {author} {\bibinfo {author} {\bibfnamefont {M}~\bibnamefont {van
  Hecke}},\ }\bibfield  {title} {\enquote {\bibinfo {title} {Jamming of soft
  particles: geometry, mechanics, scaling and isostaticity},}\ }\href@noop {}
  {\bibfield  {journal} {\bibinfo  {journal} {J. Phys.: Condens. Matter}\
  }\textbf {\bibinfo {volume} {22}},\ \bibinfo {pages} {033101} (\bibinfo
  {year} {2010})}\BibitemShut {NoStop}%
\bibitem [{\citenamefont {Cates}\ \emph {et~al.}(1998)\citenamefont {Cates},
  \citenamefont {Wittmer}, \citenamefont {Bouchaud},\ and\ \citenamefont
  {Claudin}}]{Cates1998}%
  \BibitemOpen
  \bibfield  {author} {\bibinfo {author} {\bibfnamefont {M.~E.}\ \bibnamefont
  {Cates}}, \bibinfo {author} {\bibfnamefont {J.~P.}\ \bibnamefont {Wittmer}},
  \bibinfo {author} {\bibfnamefont {J.-P.}\ \bibnamefont {Bouchaud}}, \ and\
  \bibinfo {author} {\bibfnamefont {P.}~\bibnamefont {Claudin}},\ }\bibfield
  {title} {\enquote {\bibinfo {title} {Jamming, force chains, and fragile
  matter},}\ }\href {\doibase 10.1103/PhysRevLett.81.1841} {\bibfield
  {journal} {\bibinfo  {journal} {Phys. Rev. Lett.}\ }\textbf {\bibinfo
  {volume} {81}},\ \bibinfo {pages} {1841--1844} (\bibinfo {year}
  {1998})}\BibitemShut {NoStop}%
\bibitem [{\citenamefont {Wyart}\ \emph {et~al.}(2005)\citenamefont {Wyart},
  \citenamefont {Nagel},\ and\ \citenamefont {Witten}}]{Wyart2005EPL}%
  \BibitemOpen
  \bibfield  {author} {\bibinfo {author} {\bibfnamefont {M.}~\bibnamefont
  {Wyart}}, \bibinfo {author} {\bibfnamefont {S.~R.}\ \bibnamefont {Nagel}}, \
  and\ \bibinfo {author} {\bibfnamefont {T.~A.}\ \bibnamefont {Witten}},\
  }\bibfield  {title} {\enquote {\bibinfo {title} {Geometric origin of excess
  low-frequency vibrational modes in weakly connected amorphous solids},}\
  }\href@noop {} {\bibfield  {journal} {\bibinfo  {journal} {EPL (Europhysics
  Letters)}\ }\textbf {\bibinfo {volume} {72}},\ \bibinfo {pages} {486}
  (\bibinfo {year} {2005})}\BibitemShut {NoStop}%
\bibitem [{\citenamefont {{Goldenberg}}\ and\ \citenamefont
  {{Goldhirsch}}(2005)}]{Goldenberg2005}%
  \BibitemOpen
  \bibfield  {author} {\bibinfo {author} {\bibfnamefont {C.}~\bibnamefont
  {{Goldenberg}}}\ and\ \bibinfo {author} {\bibfnamefont {I.}~\bibnamefont
  {{Goldhirsch}}},\ }\bibfield  {title} {\enquote {\bibinfo {title} {{Friction
  enhances elasticity in granular solids}},}\ }\href {\doibase
  10.1038/nature03497} {\bibfield  {journal} {\bibinfo  {journal} {\nat}\
  }\textbf {\bibinfo {volume} {435}},\ \bibinfo {pages} {188--191} (\bibinfo
  {year} {2005})}\BibitemShut {NoStop}%
\bibitem [{\citenamefont {Bouchaud}\ \emph {et~al.}(2001)\citenamefont
  {Bouchaud}, \citenamefont {Claudin}, \citenamefont {Levine},\ and\
  \citenamefont {Otto}}]{Bouchaud2001}%
  \BibitemOpen
  \bibfield  {author} {\bibinfo {author} {\bibfnamefont {J.-P.}\ \bibnamefont
  {Bouchaud}}, \bibinfo {author} {\bibfnamefont {P.}~\bibnamefont {Claudin}},
  \bibinfo {author} {\bibfnamefont {D.}~\bibnamefont {Levine}}, \ and\ \bibinfo
  {author} {\bibfnamefont {M.}~\bibnamefont {Otto}},\ }\bibfield  {title}
  {\enquote {\bibinfo {title} {Force chain splitting in granular materials: A
  mechanism for large-scale pseudo-elastic behaviour},}\ }\href {\doibase
  10.1007/s101890170100} {\bibfield  {journal} {\bibinfo  {journal} {The
  European Physical Journal E}\ }\textbf {\bibinfo {volume} {4}},\ \bibinfo
  {pages} {451--457} (\bibinfo {year} {2001})}\BibitemShut {NoStop}%
\bibitem [{\citenamefont {{Kolb}}\ \emph {et~al.}(2006)\citenamefont {{Kolb}},
  \citenamefont {{Goldenberg}}, \citenamefont {{Inagaki}},\ and\ \citenamefont
  {{Cl{\'e}ment}}}]{Kolb2006}%
  \BibitemOpen
  \bibfield  {author} {\bibinfo {author} {\bibfnamefont {E.}~\bibnamefont
  {{Kolb}}}, \bibinfo {author} {\bibfnamefont {C.}~\bibnamefont
  {{Goldenberg}}}, \bibinfo {author} {\bibfnamefont {S.}~\bibnamefont
  {{Inagaki}}}, \ and\ \bibinfo {author} {\bibfnamefont {E.}~\bibnamefont
  {{Cl{\'e}ment}}},\ }\bibfield  {title} {\enquote {\bibinfo {title}
  {{Reorganization of a two-dimensional disordered granular medium due to a
  small local cyclic perturbation}},}\ }\href {\doibase
  10.1088/1742-5468/2006/07/P07017} {\bibfield  {journal} {\bibinfo  {journal}
  {Journal of Statistical Mechanics: Theory and Experiment}\ }\textbf {\bibinfo
  {volume} {7}},\ \bibinfo {pages} {17} (\bibinfo {year} {2006})}\BibitemShut
  {NoStop}%
\bibitem [{\citenamefont {{Charbonneau}}\ \emph {et~al.}(2014)\citenamefont
  {{Charbonneau}}, \citenamefont {{Kurchan}}, \citenamefont {{Parisi}},
  \citenamefont {{Urbani}},\ and\ \citenamefont {{Zamponi}}}]{Charbonneau2014}%
  \BibitemOpen
  \bibfield  {author} {\bibinfo {author} {\bibfnamefont {P.}~\bibnamefont
  {{Charbonneau}}}, \bibinfo {author} {\bibfnamefont {J.}~\bibnamefont
  {{Kurchan}}}, \bibinfo {author} {\bibfnamefont {G.}~\bibnamefont {{Parisi}}},
  \bibinfo {author} {\bibfnamefont {P.}~\bibnamefont {{Urbani}}}, \ and\
  \bibinfo {author} {\bibfnamefont {F.}~\bibnamefont {{Zamponi}}},\ }\bibfield
  {title} {\enquote {\bibinfo {title} {{Fractal free energy landscapes in
  structural glasses}},}\ }\href {\doibase 10.1038/ncomms4725} {\bibfield
  {journal} {\bibinfo  {journal} {Nature Communications}\ }\textbf {\bibinfo
  {volume} {5}},\ \bibinfo {eid} {3725} (\bibinfo {year} {2014})},\ \Eprint
  {http://arxiv.org/abs/1404.6809} {arXiv:1404.6809 [cond-mat.dis-nn]}
  \BibitemShut {NoStop}%
\bibitem [{\citenamefont {Rainone}\ \emph {et~al.}(2015)\citenamefont
  {Rainone}, \citenamefont {Urbani}, \citenamefont {Yoshino},\ and\
  \citenamefont {Zamponi}}]{Rainone2015}%
  \BibitemOpen
  \bibfield  {author} {\bibinfo {author} {\bibfnamefont {C.}~\bibnamefont
  {Rainone}}, \bibinfo {author} {\bibfnamefont {P.}~\bibnamefont {Urbani}},
  \bibinfo {author} {\bibfnamefont {H.}~\bibnamefont {Yoshino}}, \ and\
  \bibinfo {author} {\bibfnamefont {F.}~\bibnamefont {Zamponi}},\ }\bibfield
  {title} {\enquote {\bibinfo {title} {Following the evolution of hard sphere
  glasses in infinite dimensions under external perturbations: Compression and
  shear strain},}\ }\href {\doibase 10.1103/PhysRevLett.114.015701} {\bibfield
  {journal} {\bibinfo  {journal} {Phys. Rev. Lett.}\ }\textbf {\bibinfo
  {volume} {114}},\ \bibinfo {eid} {015701} (\bibinfo {year} {2015})},\ \Eprint
  {http://arxiv.org/abs/1411.0826} {arXiv:1411.0826 [cond-mat.soft]}
  \BibitemShut {NoStop}%
\bibitem [{\citenamefont {Majmudar}\ \emph {et~al.}(2007)\citenamefont
  {Majmudar}, \citenamefont {Sperl}, \citenamefont {Luding},\ and\
  \citenamefont {Behringer}}]{Majmudar2007}%
  \BibitemOpen
  \bibfield  {author} {\bibinfo {author} {\bibfnamefont {T.~S.}\ \bibnamefont
  {Majmudar}}, \bibinfo {author} {\bibfnamefont {M.}~\bibnamefont {Sperl}},
  \bibinfo {author} {\bibfnamefont {S.}~\bibnamefont {Luding}}, \ and\ \bibinfo
  {author} {\bibfnamefont {R.~P.}\ \bibnamefont {Behringer}},\ }\bibfield
  {title} {\enquote {\bibinfo {title} {Jamming transition in granular
  systems},}\ }\href {\doibase 10.1103/PhysRevLett.98.058001} {\bibfield
  {journal} {\bibinfo  {journal} {Phys. Rev. Lett.}\ }\textbf {\bibinfo
  {volume} {98}},\ \bibinfo {pages} {058001} (\bibinfo {year}
  {2007})}\BibitemShut {NoStop}%
\bibitem [{\citenamefont {Coulais}\ \emph {et~al.}(2012)\citenamefont
  {Coulais}, \citenamefont {Behringer},\ and\ \citenamefont
  {Dauchot}}]{Coulais2012}%
  \BibitemOpen
  \bibfield  {author} {\bibinfo {author} {\bibfnamefont {C.}~\bibnamefont
  {Coulais}}, \bibinfo {author} {\bibfnamefont {R.~P.}\ \bibnamefont
  {Behringer}}, \ and\ \bibinfo {author} {\bibfnamefont {O.}~\bibnamefont
  {Dauchot}},\ }\bibfield  {title} {\enquote {\bibinfo {title} {Dynamics of the
  contacts reveals widom lines for jamming},}\ }\href@noop {} {\bibfield
  {journal} {\bibinfo  {journal} {Europhys. Lett.}\ }\textbf {\bibinfo {volume}
  {100}},\ \bibinfo {pages} {44005} (\bibinfo {year} {2012})}\BibitemShut
  {NoStop}%
\bibitem [{\citenamefont {Coulais}\ \emph
  {et~al.}(2014{\natexlab{a}})\citenamefont {Coulais}, \citenamefont
  {Behringer},\ and\ \citenamefont {Dauchot}}]{Coulais2014}%
  \BibitemOpen
  \bibfield  {author} {\bibinfo {author} {\bibfnamefont {C.}~\bibnamefont
  {Coulais}}, \bibinfo {author} {\bibfnamefont {R.~P.}\ \bibnamefont
  {Behringer}}, \ and\ \bibinfo {author} {\bibfnamefont {O.}~\bibnamefont
  {Dauchot}},\ }\bibfield  {title} {\enquote {\bibinfo {title} {How the ideal
  jamming point illuminates the world of granular media},}\ }\href {\doibase
  10.1039/C3SM51231B} {\bibfield  {journal} {\bibinfo  {journal} {Soft Matter}\
  }\textbf {\bibinfo {volume} {10}},\ \bibinfo {pages} {1519--1536} (\bibinfo
  {year} {2014}{\natexlab{a}})}\BibitemShut {NoStop}%
\bibitem [{\citenamefont {Katgert}\ and\ \citenamefont {van
  Hecke}(2010)}]{Katgert2010}%
  \BibitemOpen
  \bibfield  {author} {\bibinfo {author} {\bibfnamefont {G.}~\bibnamefont
  {Katgert}}\ and\ \bibinfo {author} {\bibfnamefont {M.}~\bibnamefont {van
  Hecke}},\ }\bibfield  {title} {\enquote {\bibinfo {title} {Jamming and
  geometry of two-dimensional foams},}\ }\href@noop {} {\bibfield  {journal}
  {\bibinfo  {journal} {Europhys. Lett.}\ }\textbf {\bibinfo {volume} {92}},\
  \bibinfo {pages} {34002} (\bibinfo {year} {2010})}\BibitemShut {NoStop}%
\bibitem [{\citenamefont {Jorjadze}\ \emph {et~al.}(2013)\citenamefont
  {Jorjadze}, \citenamefont {Pontani},\ and\ \citenamefont
  {Brujic}}]{Jorjadze2013}%
  \BibitemOpen
  \bibfield  {author} {\bibinfo {author} {\bibfnamefont {I.}~\bibnamefont
  {Jorjadze}}, \bibinfo {author} {\bibfnamefont {L.}~\bibnamefont {Pontani}}, \
  and\ \bibinfo {author} {\bibfnamefont {J.}~\bibnamefont {Brujic}},\
  }\bibfield  {title} {\enquote {\bibinfo {title} {Microscopic approach to the
  nonlinear elasticity of compressed emulsions},}\ }\href {\doibase
  10.1103/PhysRevLett.110.048302} {\bibfield  {journal} {\bibinfo  {journal}
  {Phys. Rev. Lett.}\ }\textbf {\bibinfo {volume} {110}},\ \bibinfo {pages}
  {048302} (\bibinfo {year} {2013})}\BibitemShut {NoStop}%
\bibitem [{\citenamefont {Candelier}\ and\ \citenamefont
  {Dauchot}(2010)}]{Candelier2010}%
  \BibitemOpen
  \bibfield  {author} {\bibinfo {author} {\bibfnamefont {R.}~\bibnamefont
  {Candelier}}\ and\ \bibinfo {author} {\bibfnamefont {O.}~\bibnamefont
  {Dauchot}},\ }\bibfield  {title} {\enquote {\bibinfo {title} {Journey of an
  intruder through the fluidization and jamming transitions of a dense granular
  media},}\ }\href {\doibase 10.1103/PhysRevE.81.011304} {\bibfield  {journal}
  {\bibinfo  {journal} {Phys. Rev. E}\ }\textbf {\bibinfo {volume} {81}},\
  \bibinfo {pages} {011304} (\bibinfo {year} {2010})}\BibitemShut {NoStop}%
\bibitem [{\citenamefont {Dijksman}\ \emph {et~al.}(2011)\citenamefont
  {Dijksman}, \citenamefont {Wortel}, \citenamefont {van Dellen}, \citenamefont
  {Dauchot},\ and\ \citenamefont {van Hecke}}]{Dijksman2011}%
  \BibitemOpen
  \bibfield  {author} {\bibinfo {author} {\bibfnamefont {J.~A.}\ \bibnamefont
  {Dijksman}}, \bibinfo {author} {\bibfnamefont {G.~H.}\ \bibnamefont
  {Wortel}}, \bibinfo {author} {\bibfnamefont {L.~TH}\ \bibnamefont {van
  Dellen}}, \bibinfo {author} {\bibfnamefont {O.}~\bibnamefont {Dauchot}}, \
  and\ \bibinfo {author} {\bibfnamefont {M.}~\bibnamefont {van Hecke}},\
  }\bibfield  {title} {\enquote {\bibinfo {title} {Jamming, yielding, and
  rheology of weakly vibrated granular media},}\ }\href@noop {} {\bibfield
  {journal} {\bibinfo  {journal} {Phys. Rev. Lett.}\ }\textbf {\bibinfo
  {volume} {107}},\ \bibinfo {pages} {108303} (\bibinfo {year}
  {2011})}\BibitemShut {NoStop}%
\bibitem [{\citenamefont {Katgert}\ \emph {et~al.}(2013)\citenamefont
  {Katgert}, \citenamefont {Tighe},\ and\ \citenamefont {van
  Hecke}}]{Katgert2013}%
  \BibitemOpen
  \bibfield  {author} {\bibinfo {author} {\bibfnamefont {G.}~\bibnamefont
  {Katgert}}, \bibinfo {author} {\bibfnamefont {B.~P.}\ \bibnamefont {Tighe}},
  \ and\ \bibinfo {author} {\bibfnamefont {M.}~\bibnamefont {van Hecke}},\
  }\bibfield  {title} {\enquote {\bibinfo {title} {The jamming perspective on
  wet foams},}\ }\href {\doibase 10.1039/C3SM51543E} {\bibfield  {journal}
  {\bibinfo  {journal} {Soft Matter}\ }\textbf {\bibinfo {volume} {9}},\
  \bibinfo {pages} {9739--9746} (\bibinfo {year} {2013})}\BibitemShut {NoStop}%
\bibitem [{\citenamefont {Scheffold}\ \emph {et~al.}(2013)\citenamefont
  {Scheffold}, \citenamefont {Cardinaux},\ and\ \citenamefont
  {Mason}}]{Scheffold2013}%
  \BibitemOpen
  \bibfield  {author} {\bibinfo {author} {\bibfnamefont {F}~\bibnamefont
  {Scheffold}}, \bibinfo {author} {\bibfnamefont {F}~\bibnamefont {Cardinaux}},
  \ and\ \bibinfo {author} {\bibfnamefont {T~G}\ \bibnamefont {Mason}},\
  }\bibfield  {title} {\enquote {\bibinfo {title} {Linear and nonlinear
  rheology of dense emulsions across the glass and the jamming regimes},}\
  }\href {http://stacks.iop.org/0953-8984/25/i=50/a=502101} {\bibfield
  {journal} {\bibinfo  {journal} {Journal of Physics: Condensed Matter}\
  }\textbf {\bibinfo {volume} {25}},\ \bibinfo {pages} {502101} (\bibinfo
  {year} {2013})}\BibitemShut {NoStop}%
\bibitem [{\citenamefont {Fall}\ \emph {et~al.}(2014)\citenamefont {Fall},
  \citenamefont {Weber}, \citenamefont {Pakpour}, \citenamefont {Lenoir},
  \citenamefont {Shahidzadeh}, \citenamefont {Fiscina}, \citenamefont
  {Wagner},\ and\ \citenamefont {Bonn}}]{Fall2014}%
  \BibitemOpen
  \bibfield  {author} {\bibinfo {author} {\bibfnamefont {A.}~\bibnamefont
  {Fall}}, \bibinfo {author} {\bibfnamefont {B.}~\bibnamefont {Weber}},
  \bibinfo {author} {\bibfnamefont {M.}~\bibnamefont {Pakpour}}, \bibinfo
  {author} {\bibfnamefont {N.}~\bibnamefont {Lenoir}}, \bibinfo {author}
  {\bibfnamefont {N.}~\bibnamefont {Shahidzadeh}}, \bibinfo {author}
  {\bibfnamefont {J.}~\bibnamefont {Fiscina}}, \bibinfo {author} {\bibfnamefont
  {C.}~\bibnamefont {Wagner}}, \ and\ \bibinfo {author} {\bibfnamefont
  {D.}~\bibnamefont {Bonn}},\ }\bibfield  {title} {\enquote {\bibinfo {title}
  {Sliding friction on wet and dry sand},}\ }\href {\doibase
  10.1103/PhysRevLett.112.175502} {\bibfield  {journal} {\bibinfo  {journal}
  {Phys. Rev. Lett.}\ }\textbf {\bibinfo {volume} {112}},\ \bibinfo {pages}
  {175502} (\bibinfo {year} {2014})}\BibitemShut {NoStop}%
\bibitem [{\citenamefont {Schreck}\ \emph {et~al.}(2011)\citenamefont
  {Schreck}, \citenamefont {Bertrand}, \citenamefont {O'Hern},\ and\
  \citenamefont {Shattuck}}]{Schreck2011}%
  \BibitemOpen
  \bibfield  {author} {\bibinfo {author} {\bibfnamefont {C.~F.}\ \bibnamefont
  {Schreck}}, \bibinfo {author} {\bibfnamefont {T.}~\bibnamefont {Bertrand}},
  \bibinfo {author} {\bibfnamefont {C.~S.}\ \bibnamefont {O'Hern}}, \ and\
  \bibinfo {author} {\bibfnamefont {M.~D.}\ \bibnamefont {Shattuck}},\
  }\bibfield  {title} {\enquote {\bibinfo {title} {Repulsive contact
  interactions make jammed particulate systems inherently nonharmonic},}\
  }\href {\doibase 10.1103/PhysRevLett.107.078301} {\bibfield  {journal}
  {\bibinfo  {journal} {Phys. Rev. Lett.}\ }\textbf {\bibinfo {volume} {107}},\
  \bibinfo {pages} {078301} (\bibinfo {year} {2011})}\BibitemShut {NoStop}%
\bibitem [{\citenamefont {Bertrand}\ \emph {et~al.}(2014)\citenamefont
  {Bertrand}, \citenamefont {Schreck}, \citenamefont {O'Hern},\ and\
  \citenamefont {Shattuck}}]{Bertrand2014}%
  \BibitemOpen
  \bibfield  {author} {\bibinfo {author} {\bibfnamefont {T.}~\bibnamefont
  {Bertrand}}, \bibinfo {author} {\bibfnamefont {C.~F.}\ \bibnamefont
  {Schreck}}, \bibinfo {author} {\bibfnamefont {C.~S.}\ \bibnamefont {O'Hern}},
  \ and\ \bibinfo {author} {\bibfnamefont {M.~D.}\ \bibnamefont {Shattuck}},\
  }\bibfield  {title} {\enquote {\bibinfo {title} {Hypocoordinated solids in
  particulate media},}\ }\href {\doibase 10.1103/PhysRevE.89.062203} {\bibfield
   {journal} {\bibinfo  {journal} {Phys. Rev. E}\ }\textbf {\bibinfo {volume}
  {89}},\ \bibinfo {pages} {062203} (\bibinfo {year} {2014})}\BibitemShut
  {NoStop}%
\bibitem [{\citenamefont {Goodrich}\ \emph
  {et~al.}(2014{\natexlab{a}})\citenamefont {Goodrich}, \citenamefont {Liu},\
  and\ \citenamefont {Nagel}}]{Goodrich2014}%
  \BibitemOpen
  \bibfield  {author} {\bibinfo {author} {\bibfnamefont {C.~P.}\ \bibnamefont
  {Goodrich}}, \bibinfo {author} {\bibfnamefont {A.~J.}\ \bibnamefont {Liu}}, \
  and\ \bibinfo {author} {\bibfnamefont {S.~R.}\ \bibnamefont {Nagel}},\
  }\bibfield  {title} {\enquote {\bibinfo {title} {Contact nonlinearities and
  linear response in jammed particulate packings},}\ }\href {\doibase
  10.1103/PhysRevE.90.022201} {\bibfield  {journal} {\bibinfo  {journal} {Phys.
  Rev. E}\ }\textbf {\bibinfo {volume} {90}},\ \bibinfo {pages} {022201}
  (\bibinfo {year} {2014}{\natexlab{a}})}\BibitemShut {NoStop}%
\bibitem [{\citenamefont {Lerner}\ \emph {et~al.}(2013)\citenamefont {Lerner},
  \citenamefont {During},\ and\ \citenamefont {Wyart}}]{Lerner2013}%
  \BibitemOpen
  \bibfield  {author} {\bibinfo {author} {\bibfnamefont {E.}~\bibnamefont
  {Lerner}}, \bibinfo {author} {\bibfnamefont {G.}~\bibnamefont {During}}, \
  and\ \bibinfo {author} {\bibfnamefont {M.}~\bibnamefont {Wyart}},\ }\bibfield
   {title} {\enquote {\bibinfo {title} {Low-energy non-linear excitations in
  sphere packings},}\ }\href {\doibase 10.1039/C3SM50515D} {\bibfield
  {journal} {\bibinfo  {journal} {Soft Matter}\ }\textbf {\bibinfo {volume}
  {9}},\ \bibinfo {pages} {8252--8263} (\bibinfo {year} {2013})}\BibitemShut
  {NoStop}%
\bibitem [{\citenamefont {Brito}\ \emph {et~al.}(2010)\citenamefont {Brito},
  \citenamefont {Dauchot}, \citenamefont {Biroli},\ and\ \citenamefont
  {Bouchaud}}]{Brito2010}%
  \BibitemOpen
  \bibfield  {author} {\bibinfo {author} {\bibfnamefont {C.}~\bibnamefont
  {Brito}}, \bibinfo {author} {\bibfnamefont {O.}~\bibnamefont {Dauchot}},
  \bibinfo {author} {\bibfnamefont {G.}~\bibnamefont {Biroli}}, \ and\ \bibinfo
  {author} {\bibfnamefont {J.P.}\ \bibnamefont {Bouchaud}},\ }\bibfield
  {title} {\enquote {\bibinfo {title} {Elementary excitation modes in a
  granular glass above jamming},}\ }\href {\doibase 10.1039/C001360A}
  {\bibfield  {journal} {\bibinfo  {journal} {Soft Matter}\ }\textbf {\bibinfo
  {volume} {6}},\ \bibinfo {pages} {3013--3022} (\bibinfo {year}
  {2010})}\BibitemShut {NoStop}%
\bibitem [{\citenamefont {Gomez}\ \emph {et~al.}(2012)\citenamefont {Gomez},
  \citenamefont {Turner}, \citenamefont {van Hecke}, ,\ and\ \citenamefont
  {Vitelli}}]{Gomez2012}%
  \BibitemOpen
  \bibfield  {author} {\bibinfo {author} {\bibfnamefont {L.~R.}\ \bibnamefont
  {Gomez}}, \bibinfo {author} {\bibfnamefont {A.~M.}\ \bibnamefont {Turner}},
  \bibinfo {author} {\bibfnamefont {M.}~\bibnamefont {van Hecke}}, , \ and\
  \bibinfo {author} {\bibfnamefont {V.}~\bibnamefont {Vitelli}},\ }\bibfield
  {title} {\enquote {\bibinfo {title} {Shocks near jamming},}\ }\href@noop {}
  {\bibfield  {journal} {\bibinfo  {journal} {Phys. Rev. Lett.}\ }\textbf
  {\bibinfo {volume} {108}},\ \bibinfo {pages} {058001} (\bibinfo {year}
  {2012})}\BibitemShut {NoStop}%
\bibitem [{\citenamefont {Reynolds}(1885)}]{Reynolds1885}%
  \BibitemOpen
  \bibfield  {author} {\bibinfo {author} {\bibfnamefont {O.}~\bibnamefont
  {Reynolds}},\ }\bibfield  {title} {\enquote {\bibinfo {title} {Lvii. on the
  dilatancy of media composed of rigid particles in contact. with experimental
  illustrations},}\ }\href@noop {} {\bibfield  {journal} {\bibinfo  {journal}
  {The London, Edinburgh, and Dublin Philosophical Magazine and Journal of
  Science}\ }\textbf {\bibinfo {volume} {20}},\ \bibinfo {pages} {469--481}
  (\bibinfo {year} {1885})}\BibitemShut {NoStop}%
\bibitem [{\citenamefont {Bi}\ \emph {et~al.}(2011)\citenamefont {Bi},
  \citenamefont {Zhang}, \citenamefont {Chakraborty},\ and\ \citenamefont
  {Behringer}}]{Bi2011}%
  \BibitemOpen
  \bibfield  {author} {\bibinfo {author} {\bibfnamefont {D.}~\bibnamefont
  {Bi}}, \bibinfo {author} {\bibfnamefont {J.}~\bibnamefont {Zhang}}, \bibinfo
  {author} {\bibfnamefont {B.}~\bibnamefont {Chakraborty}}, \ and\ \bibinfo
  {author} {\bibfnamefont {R.~P.}\ \bibnamefont {Behringer}},\ }\bibfield
  {title} {\enquote {\bibinfo {title} {Jamming by shear.}}\ }\href {\doibase
  10.1038/nature10667} {\bibfield  {journal} {\bibinfo  {journal} {Nature}\
  }\textbf {\bibinfo {volume} {480}},\ \bibinfo {pages} {355--358} (\bibinfo
  {year} {2011})}\BibitemShut {NoStop}%
\bibitem [{\citenamefont {Ren}\ \emph {et~al.}(2013)\citenamefont {Ren},
  \citenamefont {Dijksman},\ and\ \citenamefont {Behringer}}]{Ren2013}%
  \BibitemOpen
  \bibfield  {author} {\bibinfo {author} {\bibfnamefont {J.}~\bibnamefont
  {Ren}}, \bibinfo {author} {\bibfnamefont {J.~A.}\ \bibnamefont {Dijksman}}, \
  and\ \bibinfo {author} {\bibfnamefont {R.~P.}\ \bibnamefont {Behringer}},\
  }\bibfield  {title} {\enquote {\bibinfo {title} {Reynolds pressure and
  relaxation in a sheared granular system},}\ }\href@noop {} {\bibfield
  {journal} {\bibinfo  {journal} {Physical review letters}\ }\textbf {\bibinfo
  {volume} {110}},\ \bibinfo {pages} {018302} (\bibinfo {year}
  {2013})}\BibitemShut {NoStop}%
\bibitem [{\citenamefont {Coulais}\ \emph
  {et~al.}(2014{\natexlab{b}})\citenamefont {Coulais}, \citenamefont {Seguin},\
  and\ \citenamefont {Dauchot}}]{Coulais2014PRL}%
  \BibitemOpen
  \bibfield  {author} {\bibinfo {author} {\bibfnamefont {C.}~\bibnamefont
  {Coulais}}, \bibinfo {author} {\bibfnamefont {A.}~\bibnamefont {Seguin}}, \
  and\ \bibinfo {author} {\bibfnamefont {O.}~\bibnamefont {Dauchot}},\
  }\bibfield  {title} {\enquote {\bibinfo {title} {Shear modulus and dilatancy
  softening in granular packings above jamming},}\ }\href {\doibase
  10.1103/PhysRevLett.113.198001} {\bibfield  {journal} {\bibinfo  {journal}
  {Phys. Rev. Lett.}\ }\textbf {\bibinfo {volume} {113}},\ \bibinfo {pages}
  {198001} (\bibinfo {year} {2014}{\natexlab{b}})}\BibitemShut {NoStop}%
\bibitem [{\citenamefont {Huillard}\ \emph {et~al.}(2011)\citenamefont
  {Huillard}, \citenamefont {Noblin},\ and\ \citenamefont
  {Rajchenbach}}]{Huillard2011}%
  \BibitemOpen
  \bibfield  {author} {\bibinfo {author} {\bibfnamefont {G.}~\bibnamefont
  {Huillard}}, \bibinfo {author} {\bibfnamefont {X.}~\bibnamefont {Noblin}}, \
  and\ \bibinfo {author} {\bibfnamefont {J.}~\bibnamefont {Rajchenbach}},\
  }\bibfield  {title} {\enquote {\bibinfo {title} {Propagation of acoustic
  waves in a one-dimensional array of noncohesive cylinders},}\ }\href
  {\doibase 10.1103/PhysRevE.84.016602} {\bibfield  {journal} {\bibinfo
  {journal} {Phys. Rev. E}\ }\textbf {\bibinfo {volume} {84}},\ \bibinfo
  {pages} {016602} (\bibinfo {year} {2011})}\BibitemShut {NoStop}%
\bibitem [{\citenamefont {Johnson}(1987)}]{Johnson1987}%
  \BibitemOpen
  \bibfield  {author} {\bibinfo {author} {\bibfnamefont {K.~L.}\ \bibnamefont
  {Johnson}},\ }\href@noop {} {\emph {\bibinfo {title} {Contact mechanics}}}\
  (\bibinfo  {publisher} {Cambridge university press},\ \bibinfo {year}
  {1987})\BibitemShut {NoStop}%
\bibitem [{\citenamefont {Coulais}(2012)}]{CoulaisPhD}%
  \BibitemOpen
  \bibfield  {author} {\bibinfo {author} {\bibfnamefont {C.}~\bibnamefont
  {Coulais}},\ }\emph {\bibinfo {title} {Dense Vibrated Granular Media: From
  Stuck Liquids to Soft Solids}},\ \href@noop {} {\bibinfo {type} {Theses}},\
  \bibinfo  {school} {{Universit{\'e} Pierre et Marie Curie - Paris VI}}
  (\bibinfo {year} {2012})\BibitemShut {NoStop}%
\bibitem [{\citenamefont {Drescher}\ and\ \citenamefont {de~Josselin~de
  Jong}(1972)}]{Drescher1972}%
  \BibitemOpen
  \bibfield  {author} {\bibinfo {author} {\bibfnamefont {A.}~\bibnamefont
  {Drescher}}\ and\ \bibinfo {author} {\bibfnamefont {G.}~\bibnamefont
  {de~Josselin~de Jong}},\ }\bibfield  {title} {\enquote {\bibinfo {title}
  {Photoelastic verification of a mechanical model for the flow of a granular
  material},}\ }\href {\doibase 10.1016/0022-5096(72)90029-4} {\bibfield
  {journal} {\bibinfo  {journal} {J. Mech. Phys. Solids}\ }\textbf {\bibinfo
  {volume} {20}},\ \bibinfo {pages} {337 -- 340} (\bibinfo {year}
  {1972})}\BibitemShut {NoStop}%
\bibitem [{\citenamefont {Cundall}\ \emph {et~al.}(1982)\citenamefont
  {Cundall}, \citenamefont {Drescher},\ and\ \citenamefont
  {Strack}}]{Cundall1982}%
  \BibitemOpen
  \bibfield  {author} {\bibinfo {author} {\bibfnamefont {P.A.}\ \bibnamefont
  {Cundall}}, \bibinfo {author} {\bibfnamefont {A.}~\bibnamefont {Drescher}}, \
  and\ \bibinfo {author} {\bibfnamefont {O.D.L}\ \bibnamefont {Strack}},\
  }\href@noop {} {\bibfield  {journal} {\bibinfo  {journal} {Proc. IUTAM}\ ,\
  \bibinfo {pages} {355--370}} (\bibinfo {year} {1982})}\BibitemShut {NoStop}%
\bibitem [{\citenamefont {Cambou}\ \emph {et~al.}(2000)\citenamefont {Cambou},
  \citenamefont {Chaze},\ and\ \citenamefont {Dedecker}}]{Cambou2000}%
  \BibitemOpen
  \bibfield  {author} {\bibinfo {author} {\bibfnamefont {B.}~\bibnamefont
  {Cambou}}, \bibinfo {author} {\bibfnamefont {M.}~\bibnamefont {Chaze}}, \
  and\ \bibinfo {author} {\bibfnamefont {F.}~\bibnamefont {Dedecker}},\
  }\bibfield  {title} {\enquote {\bibinfo {title} {Change of scale in granular
  materials},}\ }\href {\doibase
  http://dx.doi.org/10.1016/S0997-7538(00)01114-1} {\bibfield  {journal}
  {\bibinfo  {journal} {European Journal of Mechanics - A/Solids}\ }\textbf
  {\bibinfo {volume} {19}},\ \bibinfo {pages} {999 -- 1014} (\bibinfo {year}
  {2000})}\BibitemShut {NoStop}%
\bibitem [{\citenamefont {Cortet}\ \emph {et~al.}(2009)\citenamefont {Cortet},
  \citenamefont {Bonamy}, \citenamefont {Daviaud}, \citenamefont {Dauchot},
  \citenamefont {Dubrulle},\ and\ \citenamefont {Renouf}}]{Cortet2009}%
  \BibitemOpen
  \bibfield  {author} {\bibinfo {author} {\bibfnamefont {P-P}\ \bibnamefont
  {Cortet}}, \bibinfo {author} {\bibfnamefont {D}~\bibnamefont {Bonamy}},
  \bibinfo {author} {\bibfnamefont {F}~\bibnamefont {Daviaud}}, \bibinfo
  {author} {\bibfnamefont {O}~\bibnamefont {Dauchot}}, \bibinfo {author}
  {\bibfnamefont {B}~\bibnamefont {Dubrulle}}, \ and\ \bibinfo {author}
  {\bibfnamefont {M}~\bibnamefont {Renouf}},\ }\bibfield  {title} {\enquote
  {\bibinfo {title} {Relevance of visco-plastic theory in a multi-directional
  inhomogeneous granular flow},}\ }\href@noop {} {\bibfield  {journal}
  {\bibinfo  {journal} {Europhys. Lett.}\ }\textbf {\bibinfo {volume} {88}},\
  \bibinfo {pages} {14001} (\bibinfo {year} {2009})}\BibitemShut {NoStop}%
\bibitem [{\citenamefont {Somfai}\ \emph {et~al.}(2007)\citenamefont {Somfai},
  \citenamefont {van Hecke}, \citenamefont {Ellenbroek}, \citenamefont
  {Shundyak},\ and\ \citenamefont {van Saarloos}}]{Somfai2007}%
  \BibitemOpen
  \bibfield  {author} {\bibinfo {author} {\bibfnamefont {E.}~\bibnamefont
  {Somfai}}, \bibinfo {author} {\bibfnamefont {M.}~\bibnamefont {van Hecke}},
  \bibinfo {author} {\bibfnamefont {W.~G.}\ \bibnamefont {Ellenbroek}},
  \bibinfo {author} {\bibfnamefont {K.}~\bibnamefont {Shundyak}}, \ and\
  \bibinfo {author} {\bibfnamefont {W.}~\bibnamefont {van Saarloos}},\
  }\bibfield  {title} {\enquote {\bibinfo {title} {Critical and noncritical
  jamming of frictional grains},}\ }\href {\doibase 10.1103/PhysRevE.75.020301}
  {\bibfield  {journal} {\bibinfo  {journal} {Phys. Rev. E}\ }\textbf {\bibinfo
  {volume} {75}},\ \bibinfo {pages} {020301} (\bibinfo {year}
  {2007})}\BibitemShut {NoStop}%
\bibitem [{\citenamefont {Goodrich}\ \emph
  {et~al.}(2014{\natexlab{b}})\citenamefont {Goodrich}, \citenamefont
  {Dagois-Bohy}, \citenamefont {Tighe}, \citenamefont {van Hecke},
  \citenamefont {Liu},\ and\ \citenamefont {Nagel}}]{Goodrich2014PRE}%
  \BibitemOpen
  \bibfield  {author} {\bibinfo {author} {\bibfnamefont {C.~P.}\ \bibnamefont
  {Goodrich}}, \bibinfo {author} {\bibfnamefont {S.}~\bibnamefont
  {Dagois-Bohy}}, \bibinfo {author} {\bibfnamefont {B.~P.}\ \bibnamefont
  {Tighe}}, \bibinfo {author} {\bibfnamefont {M.}~\bibnamefont {van Hecke}},
  \bibinfo {author} {\bibfnamefont {A.~J.}\ \bibnamefont {Liu}}, \ and\
  \bibinfo {author} {\bibfnamefont {S.~R.}\ \bibnamefont {Nagel}},\ }\bibfield
  {title} {\enquote {\bibinfo {title} {Jamming in finite systems: Stability,
  anisotropy, fluctuations, and scaling},}\ }\href {\doibase
  10.1103/PhysRevE.90.022138} {\bibfield  {journal} {\bibinfo  {journal} {Phys.
  Rev. E}\ }\textbf {\bibinfo {volume} {90}},\ \bibinfo {pages} {022138}
  (\bibinfo {year} {2014}{\natexlab{b}})}\BibitemShut {NoStop}%
\bibitem [{\citenamefont {Dagois-Bohy}\ \emph {et~al.}(2012)\citenamefont
  {Dagois-Bohy}, \citenamefont {Tighe}, \citenamefont {Simon}, \citenamefont
  {Henkes},\ and\ \citenamefont {van Hecke}}]{DagoisBohy2012}%
  \BibitemOpen
  \bibfield  {author} {\bibinfo {author} {\bibfnamefont {S.}~\bibnamefont
  {Dagois-Bohy}}, \bibinfo {author} {\bibfnamefont {B.~P.}\ \bibnamefont
  {Tighe}}, \bibinfo {author} {\bibfnamefont {J.}~\bibnamefont {Simon}},
  \bibinfo {author} {\bibfnamefont {S.}~\bibnamefont {Henkes}}, \ and\ \bibinfo
  {author} {\bibfnamefont {M.}~\bibnamefont {van Hecke}},\ }\bibfield  {title}
  {\enquote {\bibinfo {title} {Soft-sphere packings at finite pressure but
  unstable to shear},}\ }\href {\doibase 10.1103/PhysRevLett.109.095703}
  {\bibfield  {journal} {\bibinfo  {journal} {Phys. Rev. Lett.}\ }\textbf
  {\bibinfo {volume} {109}},\ \bibinfo {pages} {095703} (\bibinfo {year}
  {2012})}\BibitemShut {NoStop}%
\bibitem [{\citenamefont {Tighe}()}]{Tighe2014}%
  \BibitemOpen
  \bibfield  {author} {\bibinfo {author} {\bibfnamefont {B.}~\bibnamefont
  {Tighe}},\ }\href@noop {} {}\bibinfo {note} {Priv. comm.}\BibitemShut {Stop}%
\bibitem [{\citenamefont {Otsuki}\ and\ \citenamefont
  {Hayakawa}(2014)}]{Otsuki2014}%
  \BibitemOpen
  \bibfield  {author} {\bibinfo {author} {\bibfnamefont {M.}~\bibnamefont
  {Otsuki}}\ and\ \bibinfo {author} {\bibfnamefont {H.}~\bibnamefont
  {Hayakawa}},\ }\bibfield  {title} {\enquote {\bibinfo {title} {Avalanche
  contribution to shear modulus of granular materials},}\ }\href {\doibase
  10.1103/PhysRevE.90.042202} {\bibfield  {journal} {\bibinfo  {journal} {Phys.
  Rev. E}\ }\textbf {\bibinfo {volume} {90}},\ \bibinfo {pages} {042202}
  (\bibinfo {year} {2014})}\BibitemShut {NoStop}%
\bibitem [{\citenamefont {Tighe}(2014)}]{Tighe_granularmatter2014}%
  \BibitemOpen
  \bibfield  {author} {\bibinfo {author} {\bibfnamefont {B.~P.}\ \bibnamefont
  {Tighe}},\ }\bibfield  {title} {\enquote {\bibinfo {title} {Shear dilatancy
  in marginal solids},}\ }\href {\doibase 10.1007/s10035-013-0436-6} {\bibfield
   {journal} {\bibinfo  {journal} {Granular Matter}\ }\textbf {\bibinfo
  {volume} {16}},\ \bibinfo {pages} {203--208} (\bibinfo {year}
  {2014})}\BibitemShut {NoStop}%
\bibitem [{\citenamefont {Talamali}\ \emph {et~al.}(2008)\citenamefont
  {Talamali}, \citenamefont {Pet\"aj\"a}, \citenamefont {Vandembroucq},\ and\
  \citenamefont {Roux}}]{Talamali2008}%
  \BibitemOpen
  \bibfield  {author} {\bibinfo {author} {\bibfnamefont {M.}~\bibnamefont
  {Talamali}}, \bibinfo {author} {\bibfnamefont {V.}~\bibnamefont
  {Pet\"aj\"a}}, \bibinfo {author} {\bibfnamefont {D.}~\bibnamefont
  {Vandembroucq}}, \ and\ \bibinfo {author} {\bibfnamefont {S.}~\bibnamefont
  {Roux}},\ }\bibfield  {title} {\enquote {\bibinfo {title} {Path-independent
  integrals to identify localized plastic events in two dimensions},}\ }\href
  {\doibase 10.1103/PhysRevE.78.016109} {\bibfield  {journal} {\bibinfo
  {journal} {Phys. Rev. E}\ }\textbf {\bibinfo {volume} {78}},\ \bibinfo
  {pages} {016109} (\bibinfo {year} {2008})}\BibitemShut {NoStop}%
\bibitem [{\citenamefont {Tsamados}\ \emph {et~al.}(2009)\citenamefont
  {Tsamados}, \citenamefont {Tanguy}, \citenamefont {Goldenberg},\ and\
  \citenamefont {Barrat}}]{tsamados2009local}%
  \BibitemOpen
  \bibfield  {author} {\bibinfo {author} {\bibfnamefont {M.}~\bibnamefont
  {Tsamados}}, \bibinfo {author} {\bibfnamefont {A.}~\bibnamefont {Tanguy}},
  \bibinfo {author} {\bibfnamefont {C.}~\bibnamefont {Goldenberg}}, \ and\
  \bibinfo {author} {\bibfnamefont {J.L.}\ \bibnamefont {Barrat}},\ }\bibfield
  {title} {\enquote {\bibinfo {title} {Local elasticity map and plasticity in a
  model lennard-jones glass},}\ }\href@noop {} {\bibfield  {journal} {\bibinfo
  {journal} {Physical Review E}\ }\textbf {\bibinfo {volume} {80}},\ \bibinfo
  {pages} {026112} (\bibinfo {year} {2009})}\BibitemShut {NoStop}%
\bibitem [{\citenamefont {Cosserat}\ and\ \citenamefont
  {Cosserat}(1909)}]{Cosserat1909}%
  \BibitemOpen
  \bibfield  {author} {\bibinfo {author} {\bibfnamefont {E.}~\bibnamefont
  {Cosserat}}\ and\ \bibinfo {author} {\bibfnamefont {F.}~\bibnamefont
  {Cosserat}},\ }\href@noop {} {\emph {\bibinfo {title} {Th\'eorie des corps
  d\'eformables}}}\ (\bibinfo  {publisher} {Librairie Scientifique A. Hermann
  et Fils, Paris},\ \bibinfo {year} {1909})\BibitemShut {NoStop}%
\end{thebibliography}%
\end{document}